\documentclass[a4paper,11pt]{article}
\pdfoutput=1 

\usepackage{jheppub} 

\usepackage[T1]{fontenc} 
\usepackage{slashed}
\def\be{\begin{equation}}
\def\ee{\end{equation}}
\def\bea{\begin{array}}
\def\eea{\end{array}}
\def\beqa{\begin{eqnarray}}
\def\eeqa{\end{eqnarray}}
\def\beqas{\begin{eqnarray*}}
\def\eeqas{\end{eqnarray*}}

\def\bp{\begin{picture}}
\def\ep{\end{picture}}
\def\bc{\begin{center}}
\def\ec{\end{center}}
\def\bfig{\begin{figure}}
\def\efig{\end{figure}}

\def\bit{\begin{itemize}}
\def\eit{\end{itemize}}
\def\nn{\nonumber}
\def\f{\frac}

\def\[{\left[}
\def\]{\right]}
\def\({\left(}
\def\){\right)}

\def\..{\left.}
\def\.{\right.}
\def\tl{\tilde}
\def\ra{\rightarrow}

\def\tm{\times}

\def\da{\dagger}

\def\al{\alpha}

\def\ep{\epsilon}

\def\ga{\gamma}

\def\pr{\prime}

\allowdisplaybreaks[1]

\title{\boldmath Explanation of electron and muon $g-2$ anomalies in AMSB}

\author[a,b]{Song Li,}
\author[c]{Zhuang Li,}
\author[c]{Fei Wang,}
\author[a,b]{Jin Min Yang}

\affiliation[a]{CAS Key Laboratory of Theoretical Physics, Institute of Theoretical Physics, Chinese Academy of Sciences, Beijing 100190, P.R.China}
\affiliation[b]{School of Physics Sciences, University of Chinese Academy of Sciences,  Beijing 100049, P.R.China}
\affiliation[c]{School of Physics, Zhengzhou University, 450000, ZhengZhou, P.R.China}

\emailAdd{lisong@itp.ac.cn}
\emailAdd{lizhuang@gs.zzu.edu.cn}
\emailAdd{feiwang@zzu.edu.cn}
\emailAdd{jmyang@itp.ac.cn}

\abstract{ We propose to jointly explain the electron/muon $g-2$ anomalies in the framework of anomaly mediated SUSY breaking (AMSB) scenario. Two Yukawa deflected AMSB models are proposed and discussed in depth: one with lepton-specific interactions and the other one with messenger-matter interactions. Both models are found to be able to jointly explain the anomalies at $2 \sigma$ level by naturally realizing the preferred  parameter space with $\mu M_1,\mu M_2<0$ and very heavy left-handed smuon. 
}

\begin{document} 
\maketitle
\flushbottom

\section{Introduction}
\label{sec:intro}

Fermilab released the latest measurement result of muon anomalous magnetic moment ($g-2$) in early April last year~\cite{FNAL:gmuon}. The experimental value obtained by combining this result with the BNL E821 result~\cite{BNL:gmuon} is $4.2\sigma$ above the prediction of the standard model (SM)~\cite{Aoyama:2020ynm,RBC:2018dos}:
\begin{equation} \label{eq:delta-amu}
        \Delta a_{\mu}^{\rm{Exp-SM}} =a_\mu^{\rm Exp}-a_\mu^{\rm SM} = (2.51\pm 0.59) \times 10^{-9}.
\end{equation}
Besides, with the measurement of the fine structure constant $\alpha_{\rm em}(\text{Cs})$ by the Berkeley experiment using Cs atoms~\cite{Parker:2018vye}, the Standard Model(SM) prediction~\cite{Aoyama:2019ryr} of electron $g-2$ has a $2.4\sigma$ deviation  from the experimental value~\cite{Hanneke:2008tm}:
\begin{equation} \label{eq:delta-ae}
        \Delta a_e^{\rm{Exp-SM}} =a_e^{\rm Exp}-a_e^{\rm SM}(\text{Cs}) 
        =(-8.8\pm 3.6)\times 10^{-13}.
\end{equation}
Nonetheless, another experiment of $\alpha_{\rm em}$ using $^{87}$Rb atoms~\cite{Morel:2020dww} gives a SM prediction of $a_e$~\cite{Aoyama:2012wj} in good agreement with  $a_e^{\rm{Exp}}$. Although the deviation of electron $g-2$ is still controversial, we still should not ignore its theoretical implications for the possible forms of new physics beyond the SM. This clue opens up a lot of theoretical possibilities to go beyond the SM from a verifiable point of view. Therefore, the significance of the electron/muon $g-2$ measurements is extraordinary and the plausible anomalies have brought great encouragement to high energy physics researchers.

Electron/muon $g-2$ measurements can serve as important probes for new physics beyond the SM (for a pedagogic review of muon $g-2$, see, e.g., \cite{Li:2021bbf}). According to the dependence of a lepton $g-2$ on the new physics scale $\Lambda$~\cite{book:Jegerlehner:2017}:
\begin{equation}\label{eq:delta-alepton}
    \frac{\delta a_{\ell}}{a_{\ell}}\propto \left(\frac{m_{\ell}}{\Lambda}\right)^2,
\end{equation}
 the electron/muon $g-2$ anomalies imply an upper bound of order 2 TeV for the scale $\Lambda$~\cite{Athron:2021iuf}.  New physics of this low scale may be accessible at the current/future colliders like the LHC, such as the slepton sector in weak scale supersymmetry (SUSY).

Weak scale SUSY is the leading candidate for new physics beyond the SM and has various advantages, for example, the cancelation of the quadratic divergence of the Higgs boson and the realization of genuine gauge coupling unification by the contributions of new superpartners. If weak scale SUSY is indeed the new physics beyond the SM, it should be able to explain both anomaly and manifest itself at the near future colliders.

Various attempts have been made to explain the electron/muon $g-2$ anomalies, including the non-SUSY version ~\cite{Rose:2020nxm,Botella:2020xzf,Hernandez:2021tii,Jana:2020pxx,Keung:2021rps,Li:2020dbg,Han:2018znu,Dorsner:2020aaz,Cornella:2019uxs,Calibbi:2020emz,Escribano:2021css,Crivellin:2018qmi,CarcamoHernandez:2019ydc,Bodas:2021fsy,Chen:2020tfr,Borah:2021khc,Hue:2021zyw,Hue:2021xzl,Bigaran:2021kmn,Li:2021wzv,Panda:2022kbn,Chowdhury:2022jde} and the SUSY ones. We should list the previous explanations in the SUSY framework
\begin{itemize}
\item [(i)] For the minimal supersymmetric standard model (MSSM),  the two anomalies can be jointly explained at $2\sigma$ level~\cite{Li:2021koa} despite of the challenge in finding the parameter space. To achieve this, we either need to give up lepton flavor universality to give different soft masses for selectrons and smuons, or need to introduce lepton flavor violation (LFV). Both of these ideas can give preferred SUSY corrections to $\Delta a_{e}$ and $\Delta a_{\mu}$ with different signs. If we choose lepton flavor non-universality, the lightest neutralino $\tilde\chi_1^0$ cannot serve as dark matter candidate and some other dark matter candidates, such as the pseudo-goldstino, must be introduced. 
\item [(ii)] The assumption of 1-3 flavor violation in the MSSM
was found \cite{Dutta:2018fge} to be able to explain the electron/muon $g-2$ anomaly, which can survive the LFV constraints such as ${Br}(\mu \to e \gamma)$~\cite{TheMEG:2016wtm}. 
\item [(iii)] The effect of the SUSY threshold correction to the Yukawa couplings was found to be able to explain the electron/muon $g-2$ ~\cite{Endo:2019bcj}. 
\item [(iv)] The introduction of non-holomorphic interactions to the MSSM \cite{Ali:2021kxa} can successfully give the required contributions to the electron/muon $g-2$. 
\item [(v)] Going beyond the minimal framework, both the B-L MSSM and some extended NMSSM (next-to-minimal MSSM) were found to be able to give sufficient contributions to the electron/muon $g-2$
~\cite{Yang:2020bmh,Cao:2021lmj}. 
\end{itemize} 

In this work we retrain our study in the minimal framework of SUSY (for recent reviews see, e.g., \cite{Wang:2022rfd,Baer:2020kwz}).  We note that the studies in \cite{Li:2021koa,Badziak:2019gaf} had provided the preferred MSSM parameter space to explain the electron/muon $g-2$ anomalies with lepton flavor non-universality. The solution in ~\cite{Badziak:2019gaf} adopts $\mu M_1<0$ and the very heavy right-handed smuon; while the solution in ~\cite{Li:2021koa} adopts $\mu M_1,\mu M_2<0$ and heavy left-handed smuon. 

Although the electron/muon $g-2$ anomalies can be jointly explained in the MSSM, which notoriously has a large number of free parameters, it is still interesting to seek UV completion models that can naturally generate the previous preferred spectrum and survive all the current collider bounds.  
In this work, we try to implement the preferred parameter space in ~\cite{Li:2021koa} for electron/muon $g-2$ anomaly
within proper predictive UV completion models from SUSY breaking mechanisms, i.e., the anomaly mediation of SUSY breaking (AMSB) scenarios. 

This article is organized as follows. In Sec.~\ref{sec: modelconstrution}, we introduce our AMSB models. In Sec.~\ref{sec:scanresults}, we scan over the parameter space of the AMSB models to explain the electron/muon $g-2$. Finally, we conclude in Sec.~\ref{sec:conclusions}.

\section{Explanations of both anomalies in AMSB}
\label{sec: modelconstrution}
 To construct realistic SUSY models, it is necessary to include additional dynamics to spontaneously break SUSY. A common approach is to adopt an O'Raifeartaigh model and assume some interactions to be responsible for the mediation of the SUSY breaking effects in the hidden sector to the visible sector with ordinary matter fields. So, the SUSY breaking mechanism can act as an UV completion of low energy SUSY models and fully determines the low energy soft SUSY breaking spectrum with few free parameters. 
   Many SUSY breaking mechanisms hav been proposed in the literature, for example, the mSUGRA, the gauge mediated SUSY breaking (GMSB) mechanism, and the anomaly mediated SUSY breaking (AMSB) mechanism.  The AMSB mechanism is a very predictive one with only one free parameter in its minimal version, i.e., the gravitino mass $m_{3/2}$. However, it is known that the minimal AMSB is bothered by the tachyonic slepton problem and thus it must be extended. Among various proposals to solve such a problem, the most elegant one is the deflected AMSB, which introduces additional gauge or Yukawa couplings to deflect the AMSB trajectory so as that positive slepton square masses can be obtained.
  
   The AMSB type scenarios can be welcome to solve the muon $g-2$ anomaly. The (otherwise) negative slepton squared masses can easily be tuned to small positive values, which are preferred to give large SUSY contributions to muon $g-2$. In fact, the SUSY contributions to muon $g-2$, dominated by the chargino-sneutrino and the neutralino-smuon loops, can be estimated by~\cite{Moroi:1995yh}
 \begin{align}
 \Delta a_{\mu }(\tilde{W}, \tilde{H}, \tilde{\nu}_\mu)
 &= \frac{\alpha_2}{4\pi} \frac{m_\mu^2}{M_2 \mu} \tan\beta\cdot
f_C
 \left( \frac{M_2 ^2}{m_{\tilde{\nu }}^2}, \frac{\mu ^2}{m_{\tilde{\nu }}^2}  \right) ,
 \label{eq:WHsnu} \\
 \Delta a_{\mu }(\tilde{W}, \tilde{H},  \tilde{\mu}_L)
 &= - \frac{\alpha_2}{8\pi} \frac{m_\mu^2}{M_2 \mu} \tan\beta\cdot
 f_N
 \left( \frac{M_2 ^2}{m_{\tilde{\mu }_L}^2}, \frac{\mu ^2}{m_{\tilde{\mu }_L}^2} \right),
 \label{eq:WHmuL}  \\
 \Delta a_{\mu }(\tilde{B},\tilde{H},  \tilde{\mu }_L)
  &= \frac{\alpha_Y}{8\pi} \frac{m_\mu^2}{M_1 \mu} \tan\beta\cdot
 f_N
 \left( \frac{M_1 ^2}{m_{\tilde{\mu }_L}^2}, \frac{\mu ^2}{m_{\tilde{\mu }_L}^2} \right),
 \label{eq:BHmuL} \\
  \Delta a_{\mu }(\tilde{B}, \tilde{H},  \tilde{\mu }_R)
  &= - \frac{\alpha_Y}{4\pi} \frac{m_{\mu }^2}{M_1 \mu} \tan \beta \cdot
  f_N \left( \frac{M_1 ^2}{m_{\tilde{\mu }_R}^2}, \frac{\mu ^2}{m_{\tilde{\mu }_R}^2} \right), \label{eq:BHmuR} \\
 \Delta a_{\mu }(\tilde{\mu }_L, \tilde{\mu }_R,\tilde{B})
 &= \frac{\alpha_Y}{4\pi} \frac{m_{\mu }^2 M_1 \mu}{m_{\tilde{\mu }_L}^2 m_{\tilde{\mu }_R}^2}  \tan \beta\cdot
 f_N \left( \frac{m_{\tilde{\mu }_L}^2}{M_1^2}, \frac{m_{\tilde{\mu }_R}^2}{M_1^2}\right), \label{eq:BmuLR}
\end{align} 
at the leading order of $\tan\beta$ and $m_W/m_{SUSY}$,
with $m_\mu$ being the muon mass, $m_{SUSY}$ the SUSY-breaking masses and $\mu$ the Higgsino mass. The loop functions, defined as
\begin{align}
&f_C(x,y)= xy
\left[
\frac{5-3(x+y)+xy}{(x-1)^2(y-1)^2}
-\frac{2\log x}{(x-y)(x-1)^3}
+\frac{2\log y}{(x-y)(y-1)^3}
\right]\,,
\\
&f_N(x,y)= xy
\left[
\frac{-3+x+y+xy}{(x-1)^2(y-1)^2}
+\frac{2x\log x}{(x-y)(x-1)^3}
-\frac{2y\log y}{(x-y)(y-1)^3}
\right]\,,
\label{moroi3}
\end{align}
are monochromatically increasing for $x>0,y>0$ and satisfy $0\le f_{C,N}(x,y) \le 1$. They satisfy $f_C(1,1)=1/2$ and $f_N(1,1)=1/6$ in the limit of degenerate masses. The SUSY contributions to the muon $g-2$ will be enhanced for small soft SUSY breaking masses and large $\tan\beta$. Similar expressions can be given for electron $g-2$ after replacing properly the couplings and mass parameters for muon by those for electron.

 It is challenging to explain both anomalies in a unified framework. Without any flavor violation in the lepton sector, new physics contributions to the lepton $g-2$ are in general scaled with the corresponding lepton square mass. However, it can seen that in such a case
 \beqa
 \f{\Delta a_e}{\Delta a_\mu}\sim \f{m_e^2}{m_\mu^2}\sim 2.4\tm 10^{-5}~,
 \eeqa  
which is about 10 times smaller than the experimental results. As noted previously in the introduction, possible solutions in SUSY include large non-universal trilinear A-terms~\cite{Crivellin:2010ty} or the introduction of flavor violating off-diagonal elements in the slepton mass matrices~\cite{Dutta:2018fge} etc. Without introducing explicit flavor mixing, the two anomaly can also be explained by arranging the bino-slepton and chargino-sneutrino contributions differently
between the electron and muon sectors, which requires heavy left-hand smuon~\cite{Li:2021koa} or light selectrons,  wino and heavy higgsino~\cite{Badziak:2019gaf}.  
 
 Motivated by the previous solutions, we propose two scenarios in the AMSB framework to explain both anomalies. Both predictive scenarios can cover some of the viable regions allowed by the previous solutions \cite{Li:2021koa} or ~\cite{Badziak:2019gaf} etc at low energy. 
 
\subsection{Yukawa deflected AMSB with lepton-specific interactions}
We introduce two Higgs doublets superfields $\Phi_1({\bf 1,2,-1/2})$ and $\Phi_2({\bf 1,2,1/2})$ as well right-handed (RH) neutrino superfields to accommodate typical lepton-specific interactions involving the leptonphilic Higgs $\Phi_1$ and neutrino mass generation related terms~\footnote{The couplings of the quark sector to $\Phi_1$ and $\Phi_2$ are chosen to be negligible. Such tiny couplings are natural in SUSY because of non-renormalization theorem.}. It can be checked that the gauge anomalies are canceled. 
The two additional Higgs doublets can be fitted into ${\bf 5}$ and $\bar{5}$ representation of $SU(5)$ with proper doublet-triplet splitting mechanism to keep the triplets masses near the GUT scale.  
The superpotential takes the following form 
\beqa
W\supseteq W_{MSSM}+y_{N;ab}L_{L,a} H_u N^c_{L;b}+ y_{E;a b}L_{L,a}\Phi_1 E^c_{L;b}+\tl{\mu}\Phi_1\Phi_2+\f{M_{ab}}{2} N^c_{L;a} N^c_{L;b}~.
\eeqa
 The value of $\tl{\mu}$ are chosen to ensure that the Higgs doublets $\Phi_{1,2}$  will not acquire VEVs to break the electroweak symmetry. For simplicity, we adopt the simplest case with $y_{E;a,b}=y_{E;a}\delta_{a,b}$ and $M_{ab}=M \delta_{ab}$.

After integrating out the heavy RH neutrino, we can obtain the soft SUSY breaking parameters at the RH neutrino scale through the wavefunction renormalization approach. 
The gaugino masses can be calculated to be
\beqa\label{eq:AMSBss-binowinosoftmass}
M_i= F_\phi\f{\al_i(\mu)}{4\pi} b_i~,
\eeqa
at the RH neutrino mass scale $M$ with the corresponding beta function $(b_3,b_2,b_1)=(-3,~2,~36/5)$ and the compensator F-term VEV $F_\phi\simeq m_{3/2}$ being identified to be the gravitino mass.

The soft SUSY breaking spectrum at scale $M$ takes the following forms for trilinear couplings 
\beqa
&&
(A_t,A_b,A_\tau,A_{E;a})
=-\f{F_\phi}{16\pi^2}\left( \tl{G}_{y_t}~,
~~\tl{G}_{y_b}~,~~\tl{G}_{y_\tau}~,
~~\tl{G}_{y_{E;a}}~\right) 
\eeqa
with
\beqa
\tl{G}_{y_t}&=& 6y_t^2+y_b^2+Tr\(y_{N}^\da y_{N}\)-(\f{16}{3}g_3^2+3g_2^2+\f{13}{15}g_1^2)~,\\
\tl{G}_{y_b}&=& y_t^2+6y_b^2+y_\tau^2-(\f{16}{3}g_3^2+3g_2^2+\f{7}{15}g_1^2)~,\\
\tl{G}_{y_\tau}&=&3y_b^2+4y_\tau^2+3y_{E;3}^2+\(y_{N}^\da y_{N}\)_{33}-(3g_2^2+\f{9}{5}g_1^2)~,\\
\tl{G}_{y_{E;1}}&=&4y_{E;1}^2+y_{E;2}^2+y_{E;3}^2+\sum\limits_{c} |y_{N;1c}|^2-(3g_2^2+\f{9}{5}g_1^2)~,\\
\tl{G}_{y_{E;2}}&=&y_{E;1}^2+4y_{E;2}^2+y_{E;3}^2+\sum\limits_{c} |y_{N;2c}|^2-(3g_2^2+\f{9}{5}g_1^2)~,\\
\tl{G}_{y_{E;3}}&=&y_{E;1}^2+y_{E;2}^2+4y_{E;3}^2+3y_{\tau}^2+\sum\limits_{c} |y_{N;3c}|^2-(3g_2^2+\f{9}{5}g_1^2)~.
\eeqa
The forms for the soft SUSY breaking scalar masses are calculated to be 
\small
\beqa
&& m^2_{{H}_u}=\f{F_\phi^2}{16\pi^2}\[\f{3}{2}G_2\al^2_2+\f{3}{10}G_1\al^2_1\]
+\f{F_\phi^2}{(16\pi^2)^2}\[3y_t^2\tl{G}_{y_t}+\sum\limits_{a,b}|y_{N;ab}|^2\tl{G}_{y_{N;ab}}\],\\
&& m^2_{{H}_d}=\f{F_\phi^2}{16\pi^2}\[\f{3}{2}G_2\al^2_2+\f{3}{10}G_1\al^2_1\]
+\f{F_\phi^2}{(16\pi^2)^2}\[3y_b^2\tl{G}_{y_b}+y_\tau^2\tl{G}_{y_\tau}\]~,\\
&&m^2_{\tl{Q}_{L;1,2}}=\f{F_\phi^2}{16\pi^2}\[\f{8}{3} G_3 \al^2_3+\f{3}{2}G_2\al^2_2+\f{1}{30}G_1\al^2_1\]~,\\
&&m^2_{\tl{U}^c_{L;1,2}}=\f{F_\phi^2}{16\pi^2}\[\f{8}{3} G_3 \al^2_3+\f{8}{15}G_1\al^2_1\]~,\\
&&m^2_{\tl{D}^c_{L;1,2}}=\f{F_\phi^2}{16\pi^2}\[\f{8}{3} G_3 \al^2_3+\f{2}{15}G_1\al^2_1\]~,\eeqa\beqa
&&m^2_{\tl{L}_{L;a}}=\f{F_\phi^2}{16\pi^2}\[\f{3}{2}G_2\al_2^2+\f{3}{10}G_1\al_1^2\]
+\f{F_\phi^2}{(16\pi^2)^2}\[y_{E,a}^2\tl{G}_{y_{E,a}}+\sum\limits_{c} |y_{N;ac}|^2\tl{G}_{y_{N;ac}}\],\\
&&m^2_{\tl{E}_{L;a}^c}=\f{F_\phi^2}{16\pi^2}\f{6}{5}G_1\al_1^2
+\f{F_\phi^2}{(16\pi^2)^2}\[2y_{E,a}^2\tl{G}_{y_{E,a}}\],\\
&&m^2_{{\Phi}_1}=\f{F_\phi^2}{16\pi^2}\[\f{3}{2}G_2\al^2_2+\f{3}{10}G_1\al^2_1\]
+\f{F_\phi^2}{(16\pi^2)^2}\[\sum\limits_{a=1}^3y_{E,a}^2\tl{G}_{y_{E,a}}\]~,\\
&&m^2_{{\Phi}_2}=\f{F_\phi^2}{16\pi^2}\[\f{3}{2}G_2\al^2_2+\f{3}{10}G_1\al^2_1\]~,
\eeqa 
\normalsize
for Higgs fields and first two generation sfermions. The third generations soft SUSY breaking masses are given as
\beqa
m^2_{\tl{Q}_{L,3}}&=&m^2_{\tl{Q}_{L;1,2}}+F_\phi^2\f{1}{(16\pi^2)^2}\[y_t^2\tl{G}_{y_t}+y_b^2\tl{G}_{y_b}\]~,\\
m^2_{\tl{U}^c_{L,3}}&=&m^2_{\tl{U}^c_{L;1,2}}+F_\phi^2\f{1}{(16\pi^2)^2}\[2y_t^2\tl{G}_{y_t}\]~,\\
m^2_{\tl{D}^c_{L,3}}&=&m^2_{\tl{D}^c_{L;1,2}}+F_\phi^2\f{1}{(16\pi^2)^2}\[2y_b^2\tl{G}_{y_b}\]~,\\
m^2_{\tl{L}_{L,3}}&=&\f{F_\phi^2}{16\pi^2}\[\f{3}{2}G_2\al_2^2+\f{3}{10}G_1\al_1^2\]
+\f{F_\phi^2}{(16\pi^2)^2}\[y_{E,3}^2\tl{G}_{y_{E,3}}+y_\tau^2\tl{G}_{y_\tau}+\sum\limits_{c} |y_{N;ac}|^2\tl{G}_{y_{N;ac}}\]\nn \\ \\
m^2_{\tl{E}_{L,3}^c}&=&\f{F_\phi^2}{16\pi^2}\f{6}{5}G_1\al_1^2
+\f{F_\phi^2}{(16\pi^2)^2}\[2y_{E,3}^2\tl{G}_{y_{E,3}}+2y_\tau^2\tl{G}_{y_\tau}\]~,
\eeqa
with $G_i=-b_i$,$(b_3,b_2,b_1)=(-3,2,36/5 )$ and the beta function for Yukawa coupling $y_{N;ab}$ 
\small
\beqa
\tl{G}_{y_{N;ab}}&=&\f{1}{16\pi^2}\[\sum\limits_{c}
(y_{N;ac}^2+2|y_{N;cb}|^2) +y_{E,a}^2 +\sum\limits_{a,b}|y_{N;ab}|^2+3y_t^2-{3}g_2^2-\f{3}{5}g_1^2\]
\eeqa
\normalsize 
The inputs of this model have six free parameters
\beqa
F_\phi,~~ y_{E,1},~~~ y_{E,2},~~~y_{E,3},~~~\tl{\mu},~~~M_R.
\eeqa
All the input couplings take the values at the RH neutrino mass scale $M$. The low energy soft SUSY breaking spectrum can be obtained after renormalization group equation (RGE) evolution from RH neutrino scale $M$ to the electroweak (EW) scale. The tachyonic slepton problems should be solved so as that we can obtain relatively light sleptons.

As the type I neutrino seesaw mechanism is accommodated in the superpotential, the effective neutrino mass matrix can be obtained as
\beqa
m_\nu=-\f{v_u^2}{2}Y_N^TM_R^{-1} Y_N~,
\eeqa
which can be diagonalized by the PMNS mixing matrix
\beqa
m_{\nu}^{diag}=U_{PMNS}^T\cdot m_{\nu}\cdot U_{PMNS}~,
\eeqa
with
\small
\beqa
U_{PMNS}=\(\bea{ccc} c_{13}c_{12}&c_{13}s_{12}& s_{13}e^{-i\delta}\\
-c_{23}s_{12}-s_{23}s_{13}c_{12}e^{i\delta}&c_{23}c_{12}-s_{23}s_{13}s_{12}e^{i\delta}&s_{23}c_{13}\\
s_{23}s_{12}-c_{23}s_{13}c_{12}e^{i\delta}& -s_{23}c_{12}-c_{23}s_{13}s_{12}e^{i\delta}&c_{23}c_{13}
\eea\)\left(\bea{ccc}e^{i{\al_1/2}}&&\\&e^{i{\al_2/2}}&\\&&1\eea\right)
\eeqa
\normalsize 
The Dirac-type neutrino Yukawa couplings $Y_N$ can be defined in terms
of the physical neutrino parameters, up to an orthogonal complex matrix $R$
\beqa
y_N=\f{\sqrt{2}i}{v_u}\sqrt{M_R^{diag}}R\sqrt{m_{\nu}^{diag}}U^\da~.
\eeqa
In the special case with $R=1$, the Yukawa coupling $Y_N$ is determined by the PMNS matrix.
Requiring $y_N\sim {\cal O}(1)$, $M$ can be determined to be of order $10^{14}$ GeV.

The lepton flavor violation (LFV) processes, such as $l_i\ra l_j \ga$, will in general impose constraints on such neutrino extension SUSY models. The branch ratio $l_i\ra l_j \ga$ can be generally written as~\cite{Hirsch:2012ti,Kuno:1999jp}
\beqa
Br(l_i\ra l_j \ga)=\f{48\pi^3\al_{e}}{G_F}\(|A_L^{ij}|^2+|A_R^{ij}|^2\)Br(l_i\ra l_j \nu_i\bar{\nu}_j)~,
\eeqa
for
\beqa
A_L^{ij}\approx \f{\(m^2_{\tl{L}}\)_{ij}}{m_{SUSY}^4}~,~~A_R^{ij}\approx \f{\(m^2_{\tl{E}_L^c}\)_{ij}}{m_{SUSY}^4}~,
\eeqa
with $m_{SUSY}$ the typical SUSY mass scale and $m^2_{\tl{L}}, m^2_{\tl{E}_L^c}$ the doublet and singlet slepton soft mass matrices, respectively. It has been pointed out
that the Yukawa couplings involving $y_N$ can induce non-vanishing LFV entries in the mass matrices of the left-handed sleptons through radiative corrections~\cite{Rossi:2002zb} even with universal soft SUSY breaking terms at the GUT scale. Such off-diagonal entries of slepton mass matrices from RGE running between the GUT scale and the RH neutrino scale are proportional to $m_0^2 (y_N^\da Y_N) ln(M_{GUT}/M)$ in such SUGRA type scenarios. In our AMSB scenario, unlike neutrino seesaw mechanism extension of SUGRA-type mediation mechanism, the flavor keeping soft SUSY breaking parameters are given at the RH neutrino scale, which will lead to tiny off-diagonal entries. Therefore, the BRs of $l_i\ra l_j \ga$
will not impose important constrains in our scenario. On the other hand, the smallness of the off-diagonal entries for slepton mass matrices also indicate that the LFV proposal~\cite{Dutta:2018fge} to explain both muon and electron $g-2$ anomaly will not play an important role here. 

\subsection{Yukawa deflected AMSB with messenger-matter interactions}
We propose an alternative Yukawa deflection scenario for AMSB with typical messenger-matter interaction. For this deflected AMSB, the superpotential takes a form 
\beqa
W\supseteq W_{MSSM}+X ( \sum\limits_{i=1}^N y_P\tl{P}_i P_i+y_Q\tl{Q}^\pr {Q}^\pr+ \sum\limits_{a=1}^3 y_{X,a} Q^\pr {10}_a )+W(X)~,
\eeqa
with
\beqa
P^\pr(\bf{5})&=&L_L^\pr(1,2)_{-1/2}\oplus(3,1)_{1/3};  \\
Q^\pr(\overline{\bf 10})&=&Q_L^{c\pr}(\bar{3},2)_{-1/6}\oplus U_L^\pr({3},1)_{1/3}\oplus E_L^\pr(1,1)_{1}~.
\eeqa
 The messenger-matter interaction term in the superpotential upon the messenger scale can be written in components as 
\beqa
W^+\supseteq \( y_{X,a}^Q  Q_L^{c\pr} Q_{L,a}+y_{X,a}^U U^\pr_L U_{L,a}^c+y_{X,a}^E E^\pr_L E_{L,a}^c\)X~.
\eeqa
Below the messenger scale, the messenger superfields $\tl{P}_i,P_i,\tl{Q}^\pr, Q^\pr$ will be integrated out and will deflected the trajectory from ordinary AMSB trajectory. The superpotential for pseudo-moduli superfield $W(X)$ can be fairly generic and leads to a deflection parameter of either sign given by
\beqa
d \equiv \f{F_X}{M F_\phi}-1~.
\eeqa

After integrating out the heavy messengers, we can obtain the soft SUSY breaking spectrum at the messenger scale. 
The gaugino masses at the messenger scale are given as
\beqa\label{eq:DAMSB-binowinosoftmass}
M_i=-F_\phi\f{\al_i(\mu)}{4\pi}\(b_i-d \cdot\tl{\Delta}\)~,
\eeqa
with $(b_1~,b_2~,~b_3)=(33/5,~1,-3)$ and 
$\tl{\Delta}=3+N$.

The trilinear soft terms are given by
\beqa
A_{t,b,\tau}=-\f{F_\phi}{16\pi^2}\[(\tl{G}_{y_t}-d \cdot\Delta G_{y_t}),~(\tl{G}_{y_b}-d \cdot\Delta G_{y_b}),
~(\tl{G}_{y_\tau}- d\cdot \Delta G_{y_\tau})\]~,
\eeqa
with $\Delta G_{y_t}=|y_{X,3}^Q|^2+|y_{X,3}^U|^2$,
$\Delta G_{y_b}=|y_{X,3}^Q|^2$,  
$\Delta G_{y_\tau}=|y_{X,3}^E|^2$ and 
\beqa
\tl{G}_{y_t}&=& 6y_t^2+y_b^2-(\f{16}{3}g_3^2+3g_2^2+\f{13}{15}g_1^2)~,\\
\tl{G}_{y_b}&=& y_t^2+6y_b^2+y_\tau^2-(\f{16}{3}g_3^2+3g_2^2+\f{7}{15}g_1^2)~,\\
\tl{G}_{y_\tau}&=&3y_b^2+4y_\tau^2-(3g_2^2+\f{9}{5}g_1^2)~.
\eeqa
  The soft scalar masses are given by
\beqa
m^2_{{H}_u}~~&=&\f{F_\phi^2}{16\pi^2}\[\f{3}{2}G_2\al^2_2+\f{3}{10}G_1\al^2_1\]
+\f{F_\phi^2}{(16\pi^2)^2}\[3y_t^2\tl{G}_{y_t}\]+ \delta m^2_{{H}_u}~,\\
m^2_{{H}_d}~~&=&\f{F_\phi^2}{16\pi^2}\[\f{3}{2}G_2\al^2_2+\f{3}{10}G_1\al^2_1\]
+\f{F_\phi^2}{(16\pi^2)^2}\[3y_b^2\tl{G}_{y_b}+y_\tau^2\tl{G}_{y_\tau}\]+ \delta m^2_{{H}_d}~,\eeqa\beqa
m^2_{\tl{Q}_{L;a}}&=&\f{F_\phi^2}{16\pi^2}\[\f{8}{3} G_3 \al^2_3+\f{3}{2}G_2\al^2_2+\f{1}{30}G_1\al^2_1\]+ \delta m^2_{\tl{Q}_{L;a}}~,\\
m^2_{\tl{U}^c_{L;a}}&=&\f{F_\phi^2}{16\pi^2}\[\f{8}{3} G_3 \al^2_3+\f{8}{15}G_1\al^2_1\]+ \delta m^2_{\tl{U}^c_{L;a}}~,\\
m^2_{\tl{D}^c_{L;a}}&=&\f{F_\phi^2}{16\pi^2}\[\f{8}{3} G_3 \al^2_3+\f{2}{15}G_1\al^2_1\]+ \delta m^2_{\tl{D}^c_{L;a}}~,\\
m^2_{\tl{L}_{L;a}}&=&\f{F_\phi^2}{16\pi^2}\[\f{3}{2}G_2\al_2^2+\f{3}{10}G_1\al_1^2\]+ \delta m^2_{\tl{L}_{L;a}}~,\\
m^2_{\tl{E}_{L;a}^c}&=&\f{F_\phi^2}{16\pi^2}\f{6}{5}G_1\al_1^2+\delta m^2_{\tl{E}_{L;a}^c}~,
\eeqa
where  
\beqa  G_i=d^2\tl{\Delta}+2d\tl{\Delta}-b_i~,
\eeqa
and 
\small
\beqa
&&\delta m^2_{H_u,H_d}=-(d^2+2d) \f{F_\phi^2}{(16\pi^2)^2}
\(3y_t^2\Delta\tl{G}_{y_t},~~3y_b^2\Delta\tl{G}_{y_b}+y_\tau^2\Delta\tl{G}_{y_\tau}\),\\
&&\delta m^2_{\tl{Q}_{L;a}}=\f{d^2 F_\phi^2}{(16\pi^2)^2}\[|y_{X,a}^Q|^2\(G_{y_{X,a}^Q}^+\)\]~,(a=1,2),\\
&&\delta m^2_{\tl{Q}_{L;3}}=F_\phi^2\f{1}{(16\pi^2)^2}\[y_t^2\tl{G}_{y_t}+y_b^2\tl{G}_{y_b}\]
+\f{d^2 F_\phi^2}{(16\pi^2)^2}\[|y_{X,3}^Q|^2\(G_{y_{X,3}^Q}^+\)\]~\nn\\
&&  ~~~~~~~~~~\qquad -(d^2+2d) \f{F_\phi^2}{(16\pi^2)^2}\[y_t^2\Delta\tl{G}_{y_t}+y_b^2\Delta\tl{G}_{y_b}\]~,\\
&&\delta m^2_{\tl{U}^c_{L;a}}=
\f{d^2 F_\phi^2}{(16\pi^2)^2}\[|y_{X,a}^U|^2\(G_{y_{X,a}^U}^+\)\]~,\\
&&\delta m^2_{\tl{U}^c_{L;3}}=F_\phi^2\f{1}{(16\pi^2)^2}\[2y_t^2\tl{G}_{y_t}\]
+\f{d^2 F_\phi^2}{(16\pi^2)^2}\[|y_{X,3}^U|^2\(G_{y_{X,3}^U}^+\)\] \nn\\ 
&& ~~~~~~~~~~\qquad -(d^2+2d) \f{F_\phi^2}{(16\pi^2)^2}\[2y_t^2\Delta\tl{G}_{y_t}\],\\
&&\delta m^2_{\tl{D}^c_{L,3}}= F_\phi^2\f{1}{(16\pi^2)^2}\[2y_b^2\tl{G}_{y_b}\]-(d^2+2d) \f{F_\phi^2}{(16\pi^2)^2}\[2y_b^2\Delta\tl{G}_{y_b}\],\\
&&m^2_{\tl{L}_{L,3}}=F_\phi^2\f{1}{(16\pi^2)^2}\[y_\tau^2\tl{G}_{y_\tau}\]-(d^2+2d) \f{ F_\phi^2}{(16\pi^2)^2}\[y_\tau^2\Delta\tl{G}_{y_\tau}\],\\
&&\delta m^2_{\tl{E}_{L,a}^c}=\f{d^2 F_\phi^2}{(16\pi^2)^2}\[|y_{X,a}^E|^2\(G_{y_{X,a}^E}^+\)\]~,\\
&&\delta m^2_{\tl{E}_{L,3}^c}=F_\phi^2\f{1}{(16\pi^2)^2}2|y_\tau|^2\tl{G}_{y_\tau}+\f{d^2 F_\phi^2}{(16\pi^2)^2}|y_{X,3}^E|^2\(G_{y_{X,3}^E}^+\)
\nn\\ 
&& ~~~~~~~~~~ 
-(d^2+2d) \f{F_\phi^2}{(16\pi^2)^2}2|y_\tau|^2\Delta\tl{G}_{y_\tau},\\
&& \delta m^2_{\tl{D}^c_{L,a}}=\delta m^2_{\tl{L}_{L,a}}=0,
\eeqa
\normalsize
with
\small
\beqa
G^+_{y_{X,3}^Q}&=&y_t^2+y_b^2+|y_{X,3}^Q|^2+\sum\limits_{a=1}^3(7|y_{X,a}^Q|^2+3|y_{X,a}^U|^2
+|y_{X,a}^E|^2)~\nn\\
&&
+5N|y_P|^2
+11|y_Q|^2-2(\f{8}{3}g_3^2+\f{3}{2}g_2^2+\f{1}{30}g_1^2)~,\\
G^+_{y_{X,3}^U}&=&2y_t^2+|y_{X,3}^U|^2+\sum\limits_{a=1}^3(6|y_{X,a}^Q|^2+4|y_{X,a}^U|^2
+|y_{X,a}^E|^2)~\nn\\
&&
+5N|y_P|^2
+11|y_Q|^2-2(\f{8}{3}g_3^2+\f{8}{15}g_1^2)\\
G^+_{y_{X,3}^E}&=&2y_\tau^2+|y_{X,3}^E|^2+\sum\limits_{a=1}^3(6|y_{X,a}^Q|^2+3|y_{X,a}^U|^2
+2|y_{X,a}^E|^2)~\nn\\
&&
+5N|y_P|^2
+11|y_Q|^2-\f{12}{5}g_1^2
\eeqa
\normalsize
and
\small
\beqa
G^+_{y_{X,a}^Q}&=&|y_{X,a}^Q|^2+\sum\limits_{a=1}^3(7|y_{X,a}^Q|^2+3|y_{X,a}^U|^2
+|y_{X,a}^E|^2)~
+5N|y_P|^2
+11|y_Q|^2-2(\f{8}{3}g_3^2+\f{3}{2}g_2^2+\f{1}{30}g_1^2)~,\nn \\ \\
G^+_{y_{X,a}^U}&=&|y_{X,a}^U|^2+\sum\limits_{a=1}^3(6|y_{X,a}^Q|^2+4|y_{X,a}^U|^2
+|y_{X,a}^E|^2)
+5N|y_P|^2
+11|y_Q|^2-2(\f{8}{3}g_3^2+\f{8}{15}g_1^2)~,\\
G^+_{y_{X,a}^E}&=&|y_{X,a}^E|^2+\sum\limits_{a=1}^3(6|y_{X,a}^Q|^2+3|y_{X,a}^U|^2
+2|y_{X,a}^E|^2)
+5N|y_P|^2
+11|y_Q|^2-\f{12}{5}g_1^2~.
\eeqa
\normalsize

The previous forms of soft SUSY breaking parameters take values at the messenger scale $M_{mess}$. Those values needs to be RGE evolved from the messenger scale to the EW scale to get the low scale SUSY spectrum.

In numerical calculations, we neglect small RGE effects between the GUT scale and the messenger scale so as that one can adopt
$y_{X,a}^Q=y_{X,a}^U=y_{X,a}^E\equiv y_{X,a}$. Besides, the simplest possibility $y_P=y_Q=y_0$ is adopted.

The free parameters are reduced to the following set
\begin{equation}
F_\phi, ~M_{mess}, ~d, ~N, ~y_{X,1}, ~y_{X,2}, ~y_{X,3}, ~y_0~. 
\end{equation}
We can choose either sign for the deflection parameter 'd' in the range $ -5 \leq d\leq 5$. The Yukawa couplings satisfy $0 < y_{X,a},y_0<\sqrt{4\pi}$. $N$ can be chosen to be integers.

In this way the tachyonic slepton problem should be solved.

\section{Results of AMSB parameter scans}  
\label{sec:scanresults}

In this work, we use \textbf{SARAH 4.14.4}~\cite{SARAH4:2013tta} as model builder, then use \textbf{FlexibleSUSY 2.6.0}~\cite{FlexibleSUSY:2014yba,FlexibleSUSY:2017fvs} to calculate the mass spectrum and mixing matrices of the particles, and finally use \textbf{SPheno 4.0.4}~\cite{Porod:2003um,Porod:2011nf} to calculate the decay width of the particles. The calculation results show that the masses of supersymmetric particles get relatively large loop correction, so we use the masses with the loop corrections instead of the tree level masses to calculate the anomalous magnetic moments of the leptons. In addition to the one-loop results of anomalous magnetic moments, we also consider some high-loop corrections~\cite{Degrassi:1998es,Marchetti:2008hw,Carena:1999py}.

Note that in AMSB scenarios, the dark matter needs to be explained by other candidates like a super-weakly interacting massive particle (superWIMP)  \cite{Feng:2004gg}. For example, in AMSB the wino Lightest Ordinary Supersymmetric Particle (LOSP) will always lead to under-abundance of dark matter, which necessitates the existence of other dark matter species such as the axion. In some parameter regions, the LOSP can lead to overabundance of dark matter or even be color/electric charged. In such cases, the later decay of LOSP to axino or pseudo-goldstino LSP may provide a dark matter explanation. Moreover, in the explanation of $\delta a_{e,\,\mu}$ in \cite{Badziak:2019gaf,Li:2021koa}, the lightest neutralino $\tilde{\chi}_1^0$ cannot be a candidate for dark matter since its thermal density will be over-abundant (it must late decay to a superWIMP dark matter candidate, e.g., a light pseudo-goldstino with GeV mass~\cite{Argurio:2011hs,Dai:2021eah}.).
So, unless specified, we will not impose the dark matter constraints in our following discussions and concentrate only on the collider constraints.        
\subsection{Results of the Yukawa deflected AMSB with lepton-specific interactions}
\label{subsec:results-of-AMSBss}

In order to avoid complex high-energy scale boundary conditions, we first integrate out the right-handed neutrinos, which is equivalent to adding
\begin{equation}
    \frac{1}{2}\kappa_\nu\hat{l}\hat{H}_u\hat{l}\hat{H}_u
\end{equation}
into the Lagrangian, with $\kappa_\nu=y_N^{\rm T}y_N/M_R$. At the same time, we ignore the CP phase in the PMNS matrix since our calculation is not sensitive to it.  According to ~\cite{deSalas:2020pgw}, the lightest neutrino mass will be below 0.01 eV. Since $\Delta m_{21}^2\ll|\Delta m_{31}^2|\sim 2.5\times 10^{-3}\text{eV}^2$, the mass of the heaviest neutrino will be mainly determined by $\Delta m_{31}^2$. This suggests that the anomalous magnetic moment is insensitive to the neutrino mass ordering and the lightest neutrino mass. Therefore, we let the neutrino masses be normal ordering and take $m_{\nu_1}=10^{-3}\text{eV}$ for simplicity.

According to eq.~\eqref{eq:AMSBss-binowinosoftmass}, the mass parameters $M_1$ and $M_2$ take the same sign. Therefore, it is expected to satisfy the parameter requirements for the joint explanation of electron/muon $g-2$ under the condition of $\mu<0$ \cite{Li:2021koa}. On the other hand, the phases that can affect the physical results are ${\rm Arg}(\mu M_i)$ and ${\rm Arg}(\mu A_f)$~\cite{Kraml:2007pr}, so we can also take $\mu$ positive and multiply the values of $M_i$ and $A_f$ by $e^{i\pi}$,  which is equivalent to a redefinition of the phase for the corresponding fields and will not  change the physical results. 

\begin{figure*}[tbp]
    \centering 
    \includegraphics[width=.7\textwidth]{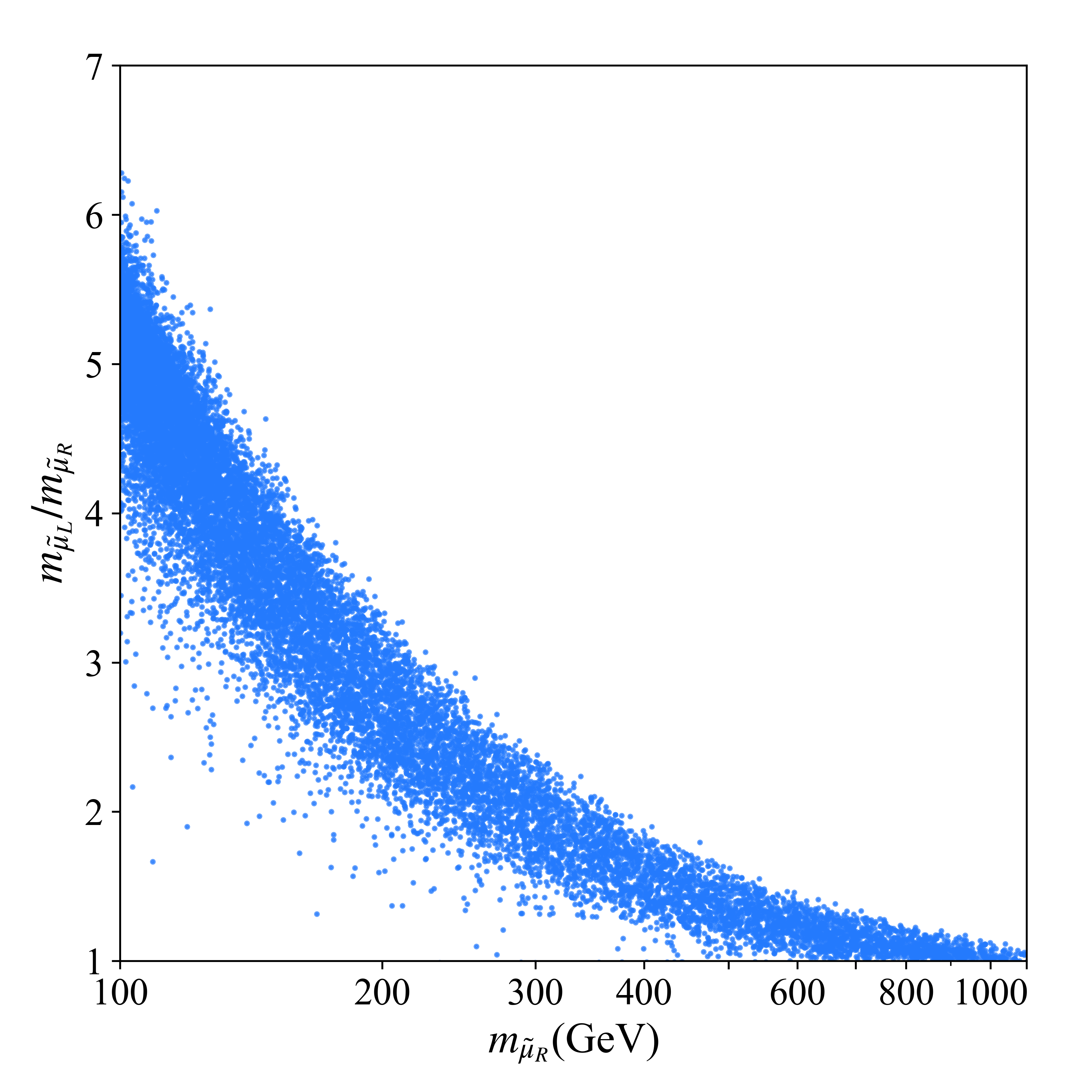}
    \caption{\label{fig:AMSBss-mass-ratio} 
    Scatter plots in the Yukawa deflected AMSB with lepton-specific interactions,
    showing a large mass ratio $m_{\tilde{\mu}_L}/m_{\tilde{\mu}_R}$.}
\end{figure*}

The scheme in \cite{Li:2021koa}  to jointly explain electron/muon $g-2$ 
requires the left-handed smuon to be significantly heavier than the right-handed smuon. Fig.~\ref{fig:AMSBss-mass-ratio} shows the scatter plots from our scan, which shows the mass ratio $m_{\tilde{\mu}_L}/m_{\tilde{\mu}_R}$ versus $m_{\tilde{\mu}_R}$. We see that in the scanned parameter region with the experimental lower bound
of 100 GeV,  the ratio $m_{\tilde{\mu}_L}/m_{\tilde{\mu}_R}$ can indeed  take a large value.   Such a large ratio of the smuon masses is achieved via the negative loop correction of $m_{\tilde{\mu}_R}$, while the tree-level mass ratio cannot be that large. Note that in this case the loop correction is sizable and higher order corrections might be needed. 

\begin{figure*}[tbp]
\centering 
\includegraphics[width=.7\textwidth]{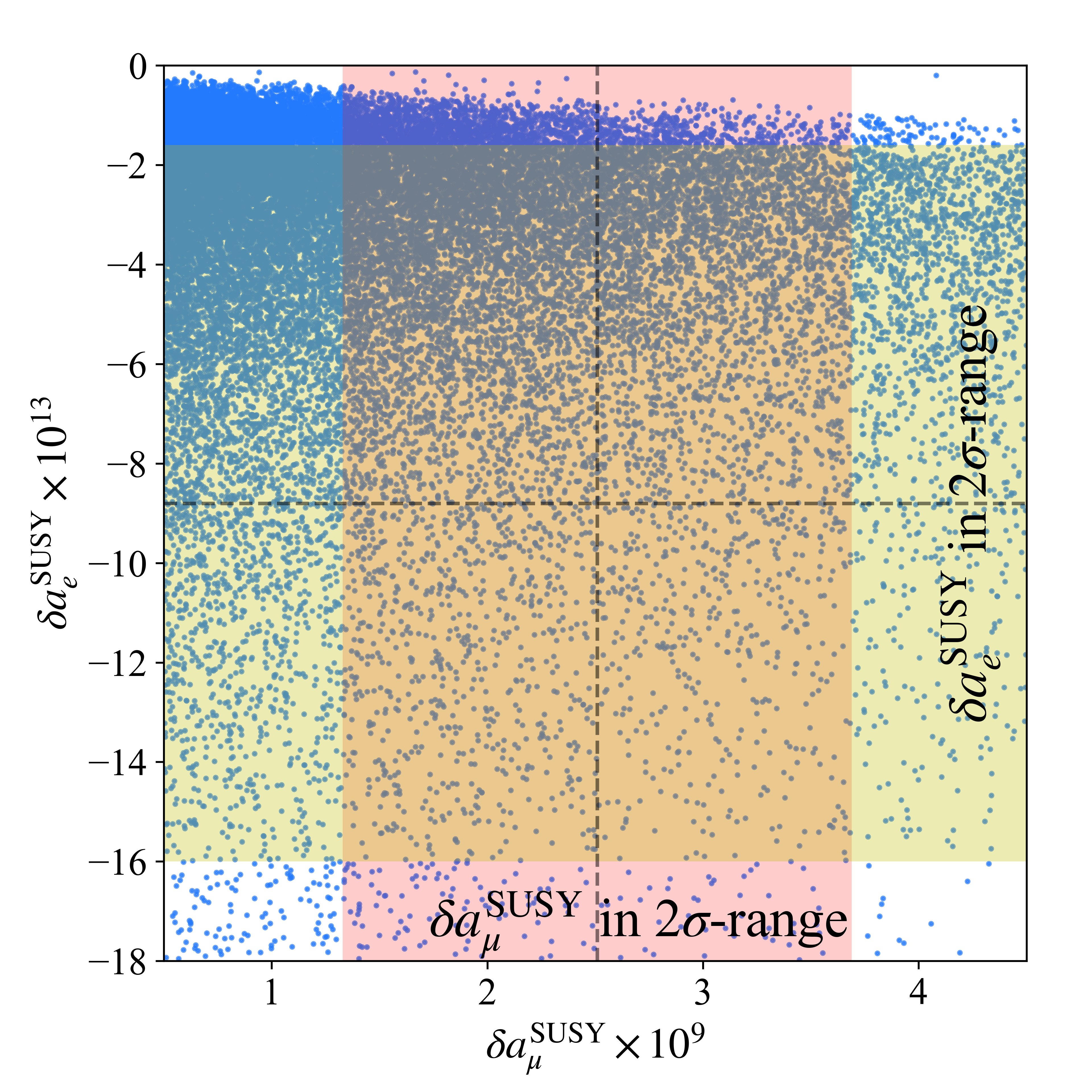}
\caption{\label{fig:AMSBss-a-ele-vs-a-mu} 
Scatter plots in the Yukawa deflected AMSB with lepton-specific interactions, showing $\delta a_e^{\rm SUSY}$ versus $\delta a_\mu^{\rm SUSY}$. The red area is the $2\sigma$-range of $\Delta a_\mu$, and the yellow area is the $2\sigma$-range of $\Delta a_e$. The gray dashed lines correspond to the central values of $\Delta a_\mu$ and $\Delta a_e$, respectively.}
\end{figure*}

Fig.~\ref{fig:AMSBss-a-ele-vs-a-mu} shows a scatter plot of $\delta a_e^{\rm SUSY}$ versus $\delta a_\mu^{\rm SUSY}$. This figure shows that it is feasible to explain the $g-2$ of electron and muon at $2\sigma$ level, albeit most samples give rather small contributions in magnitude. For the joint explanation at $2\sigma$ level, the left-handed smuon must be much heavier than the right-handed smuon, and the right-handed smuon and left-handed selectron must be lighter than 200 GeV. The reduction of $m_{\tilde{\mu}_R}$ mainly relies on the loop correction. If the magnitude of the loop correction is too large, it may lead to the tachyonic smuon. Therefore, $\delta a_\mu$ and $|\delta a_e|$ tend to take relatively small values, and they are sensitive to the relevant parameter values.

\begin{figure*}[tbp]
\centering 
\includegraphics[width=.45\textwidth]{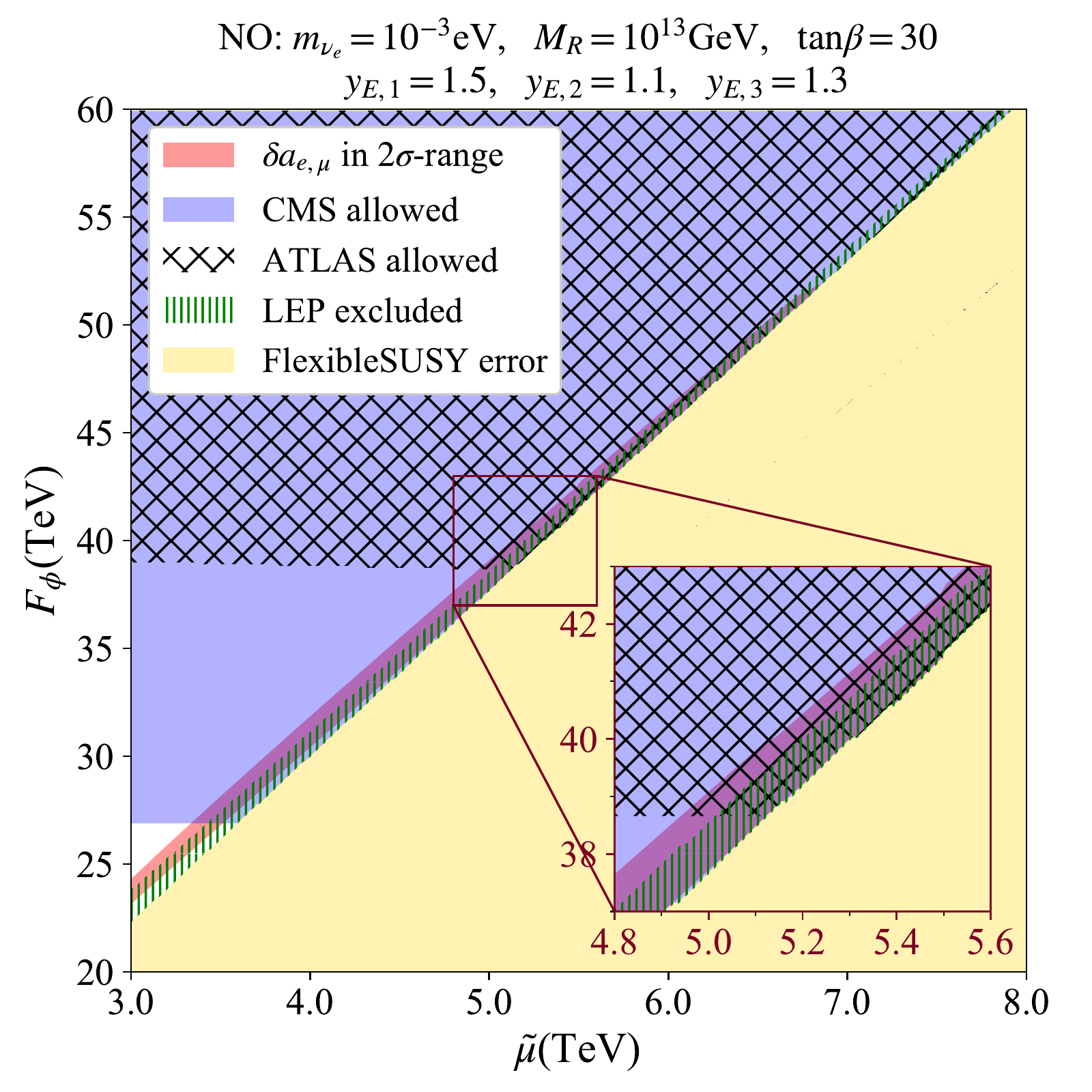}
\includegraphics[width=.45\textwidth]{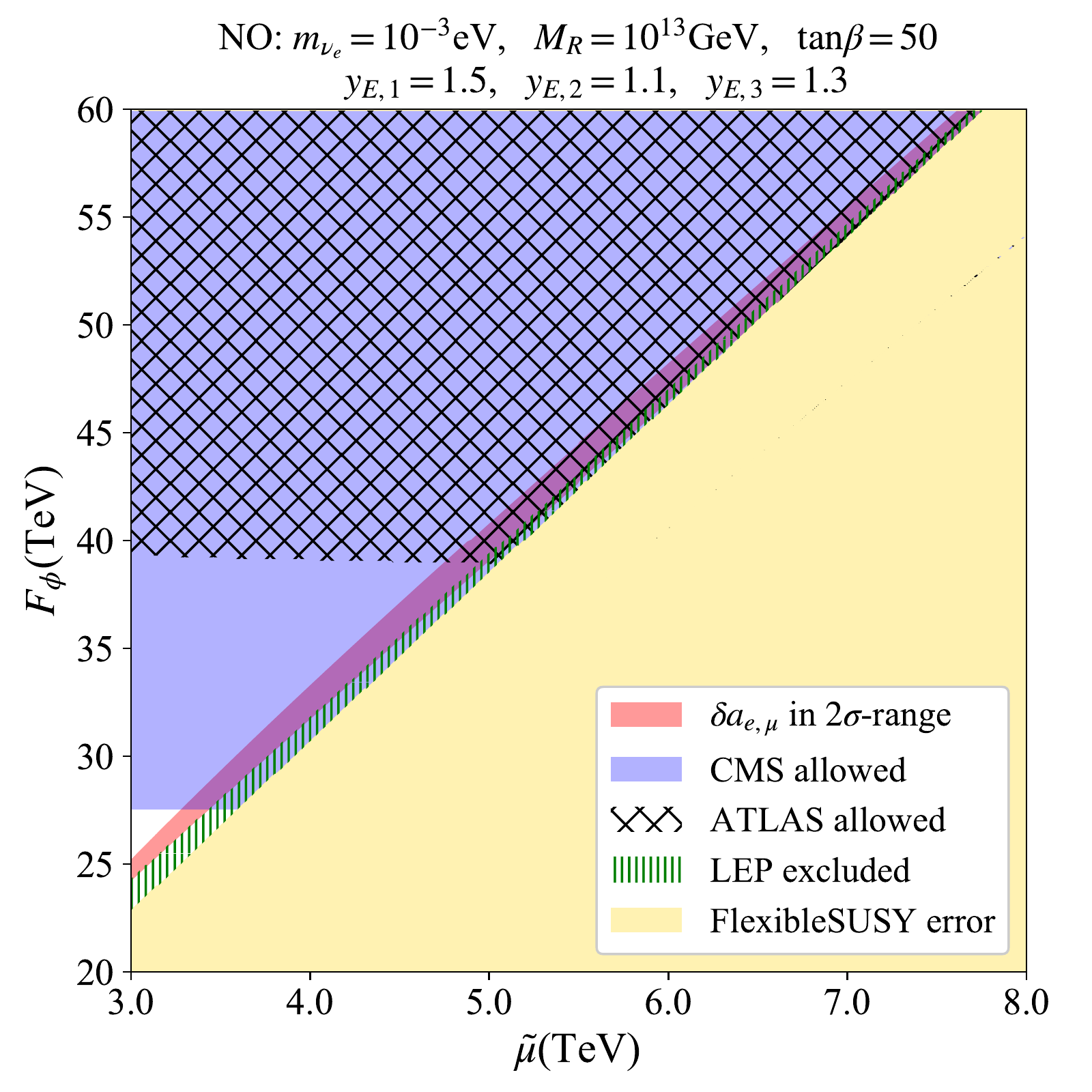}
\caption{\label{fig:AMSBss-Fphi-vs-MupIN} 
Anomalous magnetic moments and collider constraints in the  Yukawa deflected AMSB with lepton-specific interactions, shown on the plane of $F_\phi$ versus $\tilde{\mu}$. The left panel corresponds to $\tan\beta=30$ and the right panel corresponds to $\tan\beta=50$.} 
\end{figure*}

In Fig.~\ref{fig:AMSBss-a-ele-vs-a-mu} the constraints from the colliders have not been considered. If the collider constraints (especially the constraints of the LEP) are added, the range of parameters that can explain $\Delta a_{e,\mu}$ will be further narrowed. To examine the collider constraints on the parameter range, we mesh and scan the parameter space with the constraints of LEP~\cite{LEP:sleptons}, CMS~\cite{CMS:2017kxn,CMS:2020wxd,CMS:2021cox} and ATLAS~\cite{ATLAS:2019lff,ATLAS:2019lng,ATLAS:2019wgx}. The anomalous magnetic moments and collider constraints on the plane of  $|F_\phi|$ versus $\tilde{\mu}$ are shown in Fig.~\ref{fig:AMSBss-Fphi-vs-MupIN}. The regions marked in yellow indicate that the package FlexibleSUSY report errors. The main causes of such reported errors include tachyonic sfermions, non-perturbative running of parameters, absence of the electroweak phase transition and dis-convergence of the iteration. These errors all come from incorrect settings of parameters. So we just ignore these yellow un-physical regions. As shown in the figure, in order to explain $\Delta a_{e,\mu}$, the parameters $F_\phi$ and $\tilde{\mu}$ must satisfy an approximate linear relation. By this linear relationship, the the right-handed smuon mass gets a negative loop correction, so that the values of $\delta a_{e,\mu}$ can be in the $2\sigma$-range. Because the right-handed smuon mass needs to be reduced by the negative loop correction, it may be easily lower than 100 GeV which is not allowed by LEP. For a fixed $\tilde{\mu}$, if $F_\phi$ is too small, $m_{\tilde{\chi}_1^0}$ will decrease rapidly while $m_{\tilde{e}}$  will remain almost unchanged, which is  excluded by ATLAS~\cite{ATLAS:2019lff}. Compared with $\tan\beta=30$, the case of $\tan\beta=50$ can give more samples satisfying the requirements of $\delta a_{e,\mu}$. 

\begin{figure*}[tbp]
	\centering 
	\includegraphics[width=.45\textwidth]{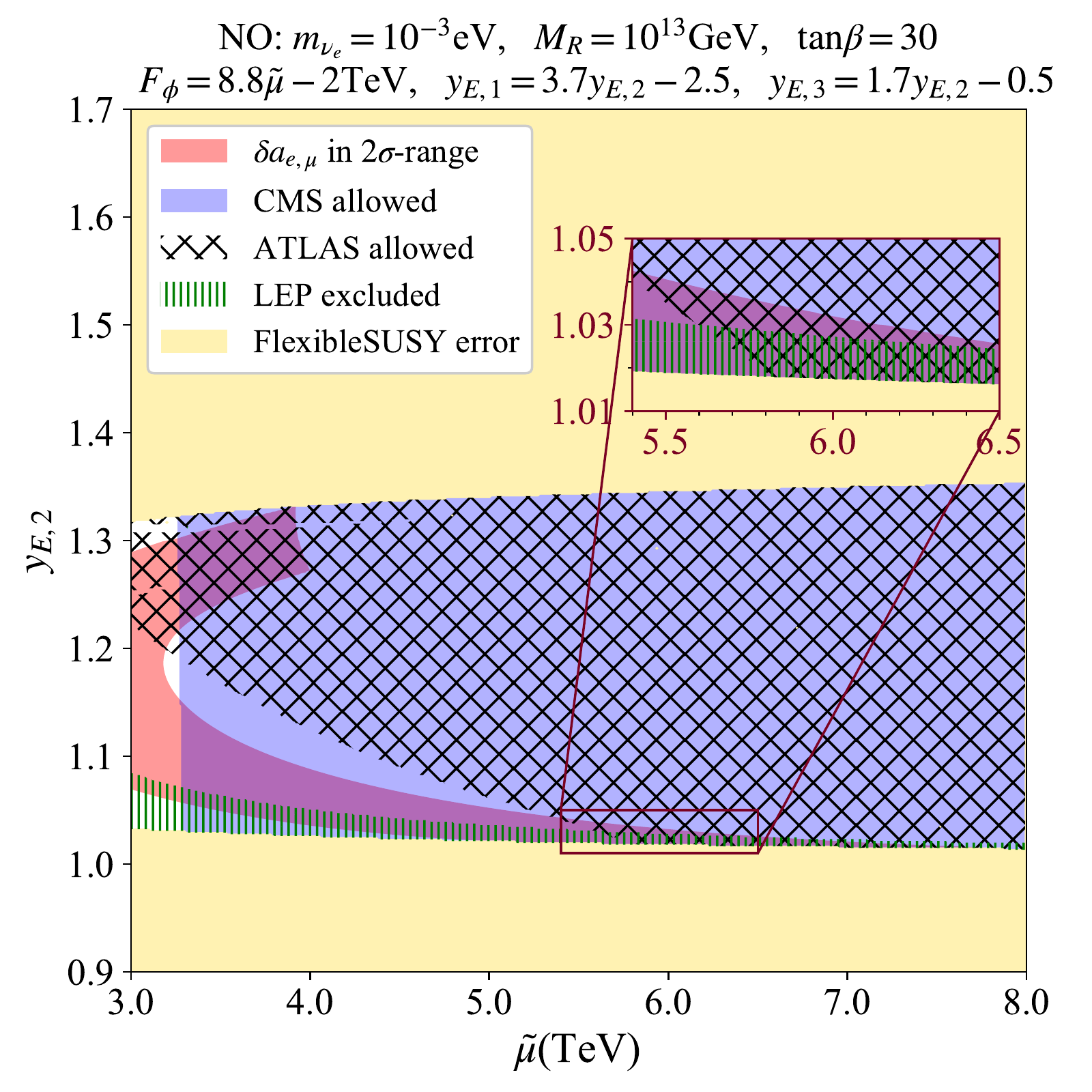}
	\includegraphics[width=.45\textwidth]{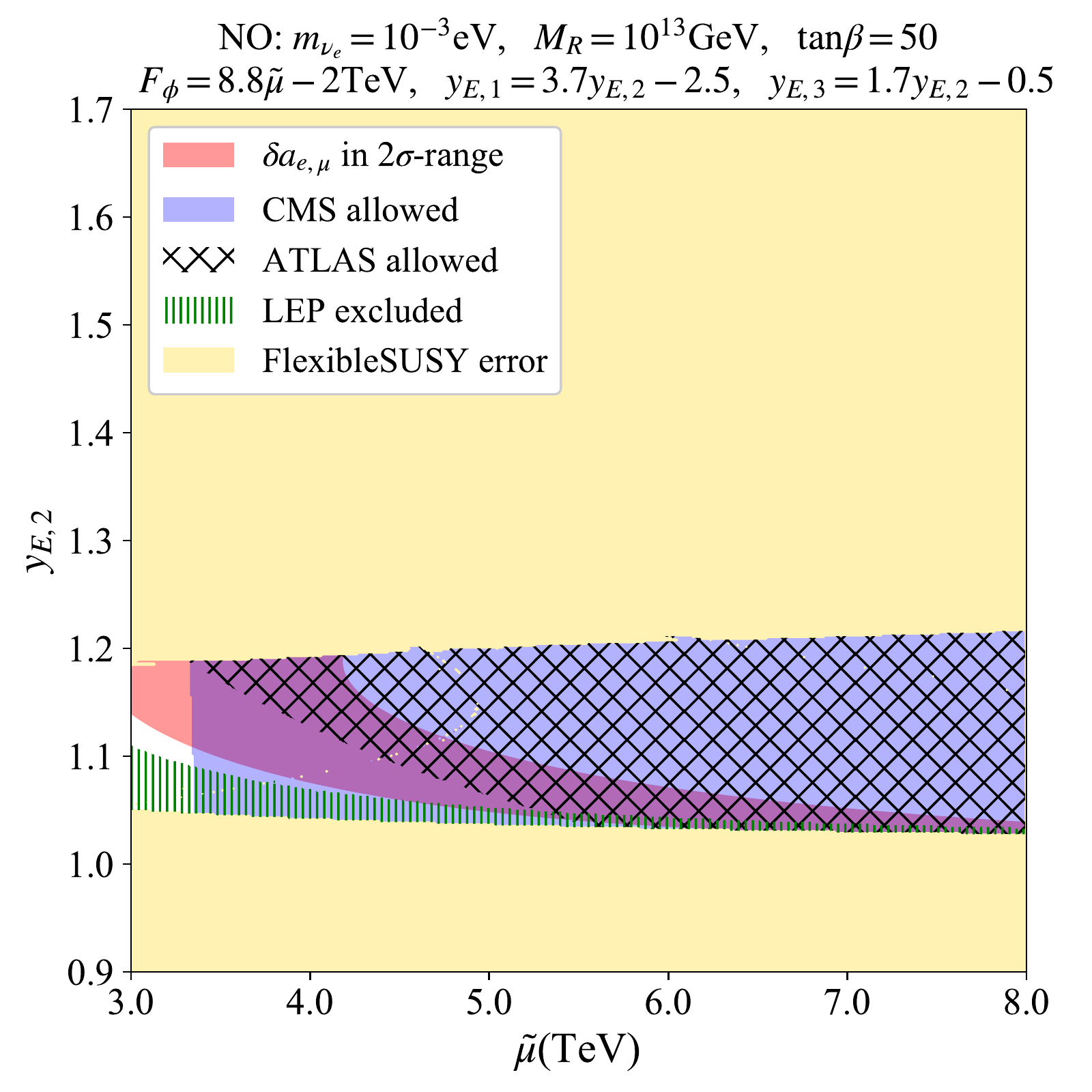}
	\caption{   \label{fig:AMSBss-Yep22-vs-MupIN} 
	Anomalous magnetic moments and collider constraints in the Yukawa deflected AMSB with lepton-specific interactions, shown on the plane of $y_{E,2}$ versus $\tilde{\mu}$. The left and right panels corresponds to $\tan\beta=30$ and  $\tan\beta=50$, respectively. Here, the approximate linear relations $F_{\phi}\simeq 8.8\tilde{\mu}-2\text{TeV}$, $y_{E,1}\simeq 3.7y_{E,2}-2.5$ and $y_{E,3}\simeq 1.7y_{E,2}-0.5$ can be estimated from our scan results.}
\end{figure*}

Fig.~\ref{fig:AMSBss-Yep22-vs-MupIN} shows the parameter $y_{E,2}$ versus $\tilde{\mu}$ under $g-2$ and collider constraints. Here the approximate linear relations $F_{\phi}\simeq 8.8\tilde{\mu}-2\text{TeV}$, $y_{E,1}\simeq 3.7y_{E,2}-2.5$ and $y_{E,3 } \simeq 1.7y_{E,2}-0.5$ can be estimated based on our scan results.  Even without considering the anomalous magnetic moment requirement, the parameter $y_{E,2}$ is constrained to a small range. When $y_{E,2}\gtrsim 1.3$ (for $\tan\beta=30$) or $y_{E,2}\gtrsim 1.2$ (for $\tan\beta=50$), the perturbative calculation will fail over a certain energy scale, and cause \textbf{FlexibleSUSY} to report an error. When $y_{E,2}\lesssim 1.04$, it will make sleptons too light and even lead to tachyonic sleptons. When the value of $\tilde{\mu}$ is relatively small and $y_{E,2}$ is too small, $m_{\tilde{\chi}_1^0}$ will become smaller, which is limited by ATLAS after combining with the masses of other sparticles. After considering the requirement of $a_{e,\mu}$, the value of $y_{E,2}$ will be restricted to a narrower range. This is mainly due to the fact that the correction of the right-handed smuon mass is sensitive to $y_{E,2}$ under the requirement of $a_{e,\mu}$.  Such a sensitive parameter dependence is a 
feature and a possible challenge for this model.

\begin{figure*}[tbp]
	\centering 
	\includegraphics[width=.45\textwidth]{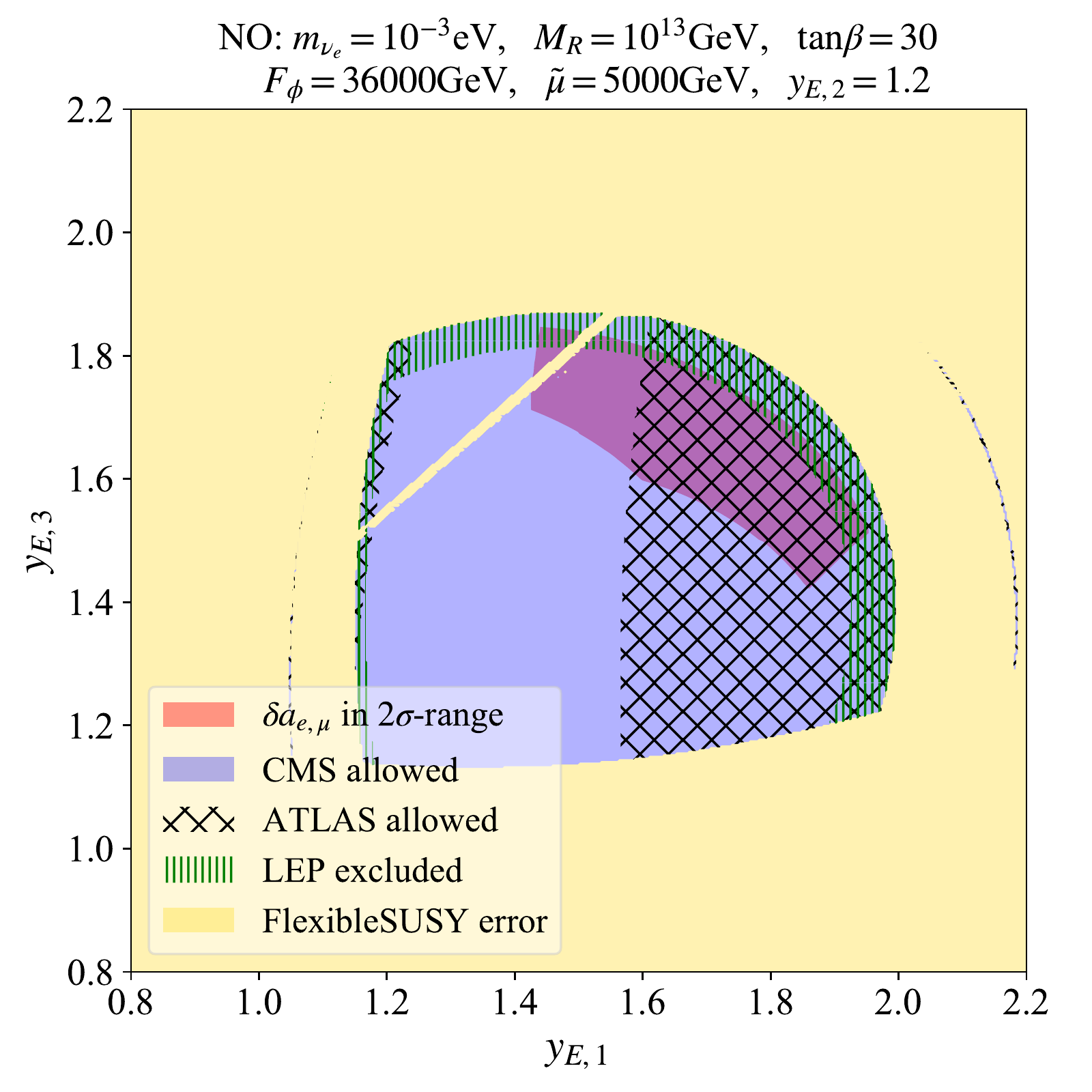}
	\includegraphics[width=.45\textwidth]{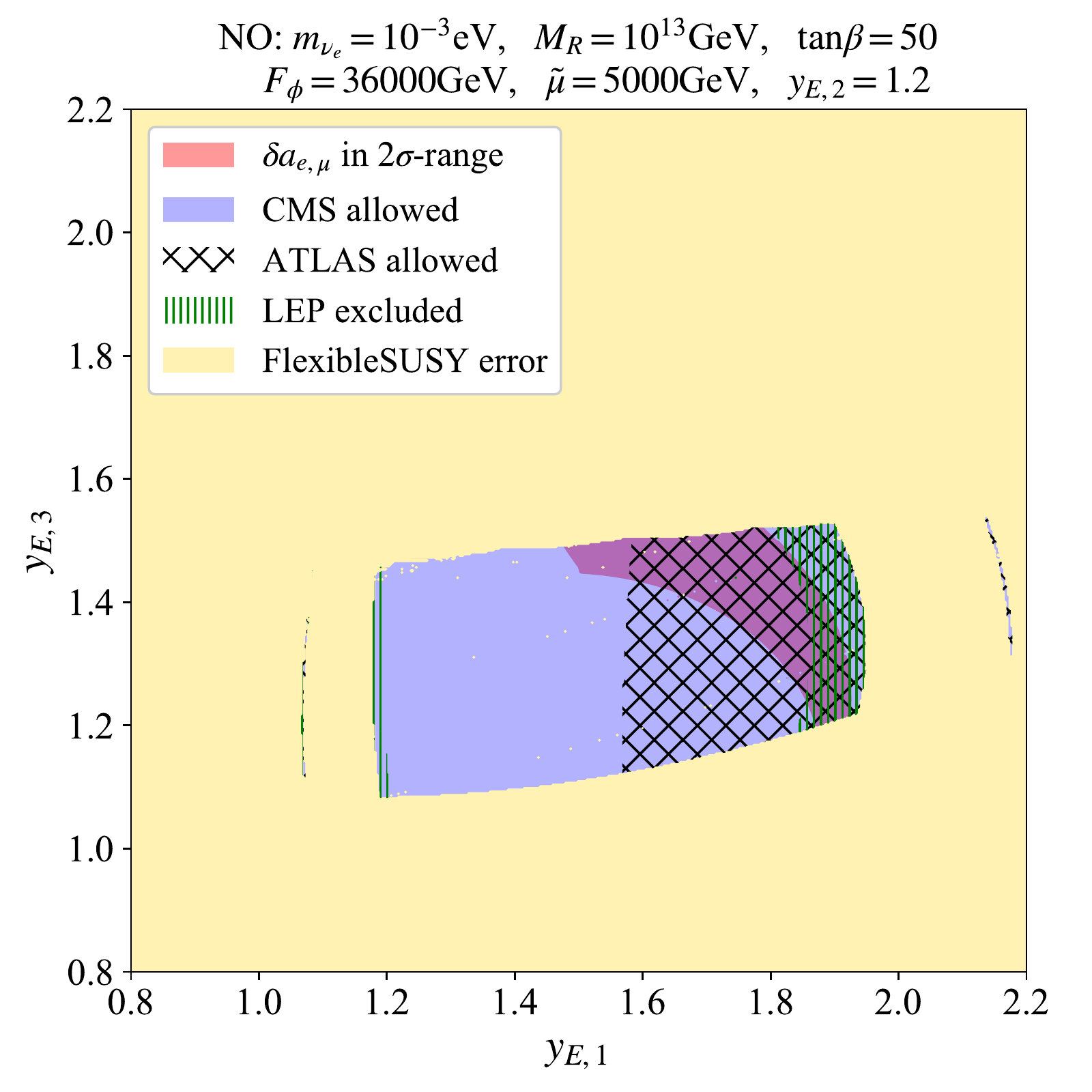}
	\caption{\label{fig:AMSBss-Yep33-vs-Yep11} 
	Anomalous magnetic moments and collider constraints in the Yukawa deflected AMSB with lepton-specific interactions, shown on the $y_{E,3}$ vs $y_{E,1}$ plane.} 
\end{figure*}

Fig.~\ref{fig:AMSBss-Yep33-vs-Yep11} shows the SUSY contributions to the anomalous magnetic moments and collider constraints on the $y_{E,3}$-$y_{E,1}$ plane. Whether $y_{E,1}$ is too large or too small, some sleptons will be constrained stringently by LEP. When $y_{E,1}$ is too small, the selectron mass will be too small, and when $y_{E,1}$ is too large, the magnitude of the loop correction for the right-handed smuon mass is too large and makes its mass too small. On the other hand, if $y_{E,1}\lesssim 1.6$, the mass $m_{\tilde{\chi}_1^0}$ will be too small to survive the bounds by ATLAS. Discussions for $y_{E,3}$ are similar to that of $y_{E,1}$, except that the stau mass is not related to $a_{e,\mu}$, so that the LEP bounds for stau are not shown in the figures. Unlike the case in Fig.~\ref{fig:AMSBss-Yep22-vs-MupIN}, in the $y_{E,1}$-$y_{E,3}$ plane,  smaller $\tan\beta$ tends to give more survived parameter points.

\subsection{Results of the Yukawa deflected AMSB with messenger-matter interactions}

For simplicity, we take $N=1$ to give  $\tilde\Delta=4$ in Eq.\eqref{eq:DAMSB-binowinosoftmass}. As discussed in \ref{subsec:results-of-AMSBss}, $\mu$ can be chosen positive, which is equivalent to $\mu<0$ with minus sign for EW gaugino masses after redefinition of their phases. We have the following three cases:
\begin{align}
	&\text{Case 1: }~~~~-5<d<\frac14:\quad b_1-d\tilde{\Delta}>0,\,b_2-d\tilde{\Delta}>0;\\
	&\text{Case 2: }~~~~\frac14<d<\frac{33}{20}:\quad b_1-d\tilde{\Delta}>0,\,b_2-d\tilde{\Delta}<0;\\
	&\text{Case 3: }~~~~\frac{33}{20}<d<5:\quad b_1-d\tilde{\Delta}<0,\,b_2-d\tilde{\Delta}<0.
\end{align}
For Case 1 and Case 3, we can get $\mu M_1,\mu M_2<0$; for Case 2, we get $\mu M_1<0,\mu M_2>0$. After scanning the parameter space in these cases, we find in  Case 2 that, most samples have too light sleptons to survive the LEP bounds, although there are parameter samples that can meet the requirement of electron/muon $g-2$. Case 3 is similar to that of Case 1. Therefore, we only concentrate on the numerical results of Case 1 in our following discussions.

\begin{figure*}[tbp]
	\centering 
	\includegraphics[width=.7\textwidth]{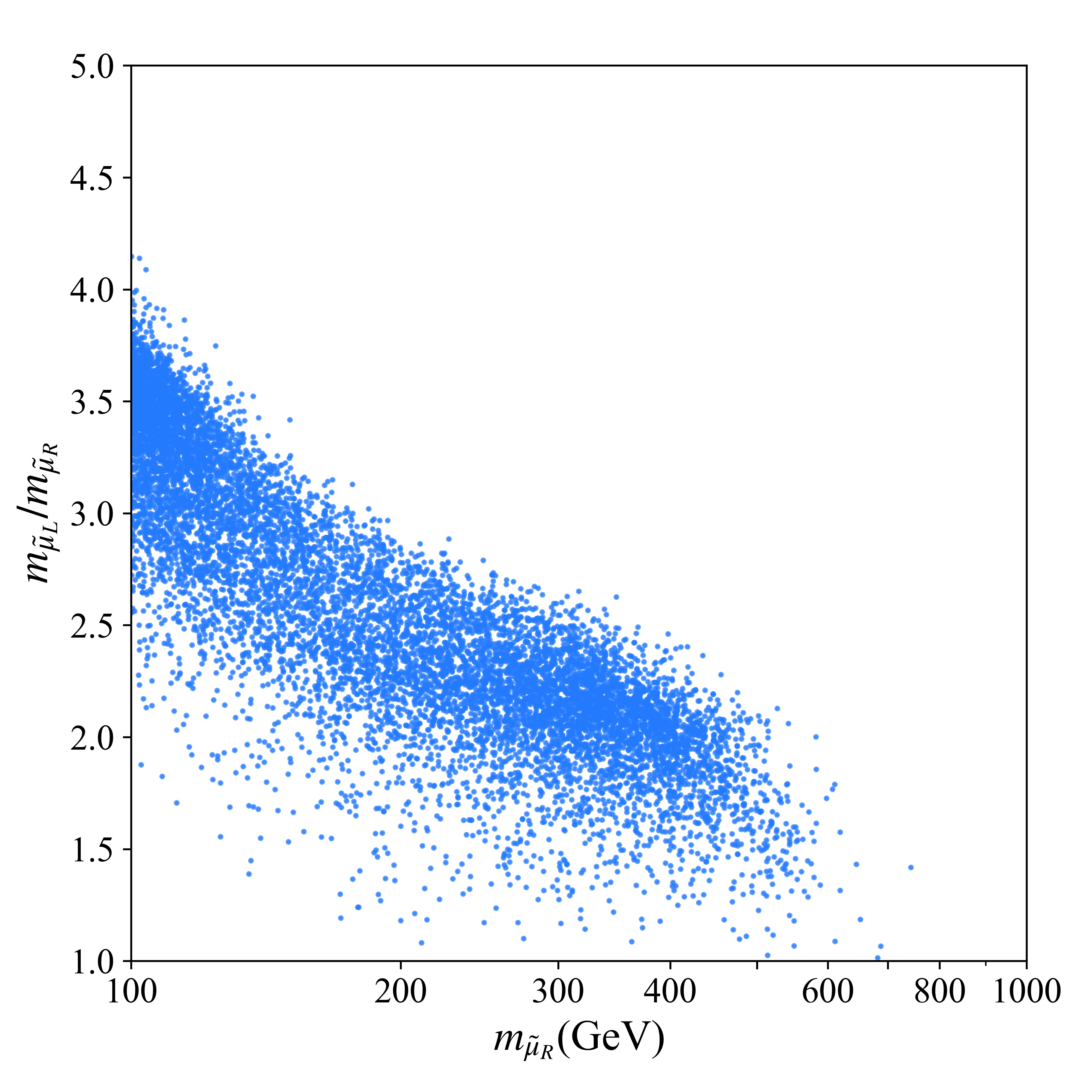}
	\caption{\label{fig:DAMSB-mass-ratio} 
	Scatter plots for the Yukawa deflected AMSB with messenger-matter interactions, showing a large mass ratio $m_{\tilde{\mu}_L}/m_{\tilde{\mu}_R}$.}
\end{figure*}

\begin{figure*}[tbp]
	\centering 
	\includegraphics[width=.6\textwidth]{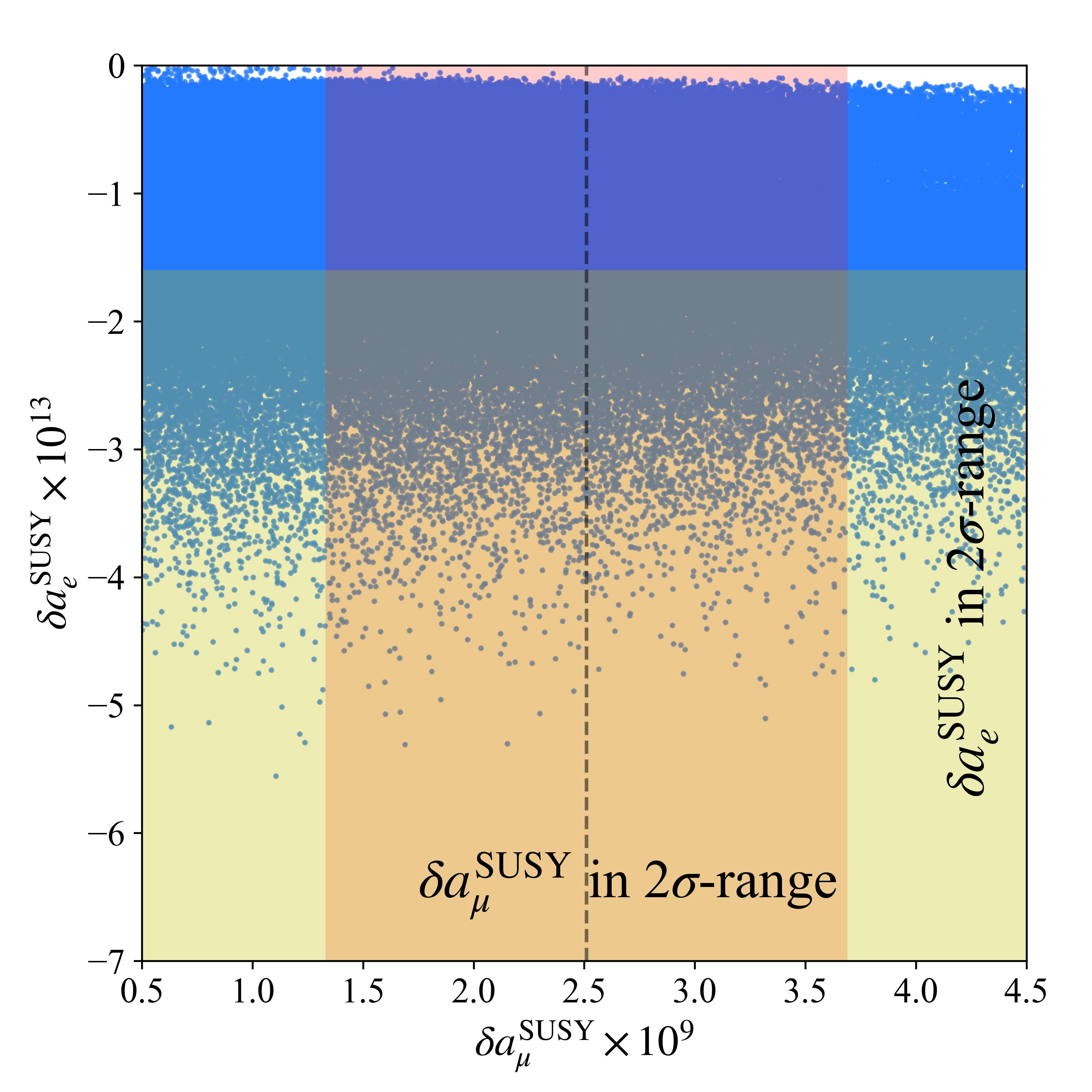}
	\caption{\label{fig:DAMSB-a-ele-vs-a-mu} 
	Scatter plots in the Yukawa deflected AMSB with messenger-matter interactions, showing  $\delta a_e^{\rm SUSY}$ and $\delta a_\mu^{\rm SUSY}$. The red area is the $2\sigma$ range of $\Delta a_\mu$, and the yellow area is the $2\sigma$ range of $\Delta a_e$. The gray dashed line corresponds to the central value of $\Delta a_\mu$.}
\end{figure*}

Fig~\ref{fig:DAMSB-mass-ratio} shows that a large mass ratio $m_{\tilde{\mu}_L}/m_{\tilde{\mu}_R}$ can be obtained in Case 1. However, from our scan we find that the ratio $m_{\tilde{\mu}_L}/m_{\tilde{\mu}_R}$ can maximally take about 4.0. In this case, $\delta a_{e}^{\rm SUSY}$ can be within the $2\sigma$ region but outside the $1\sigma$ range. When $d<1/4$, the selectron mass is relatively heavy in those parameter regions that can explain the electron and muon $g-2$, and thus the SUSY contributions to $|\delta a_{e}^{\rm SUSY}|$ is relatively suppressed, as shown in  Fig. ~\ref{fig:DAMSB-a-ele-vs-a-mu}. In order to explain the two anomalous magnetic moments simultaneously, the mass of the right-handed smuon also needs to obtain a large negative loop correction and thus is constrained by the LEP bounds on slepton masses. 

\begin{figure*}[tbp]
	\centering 
	\includegraphics[width=.45\textwidth]{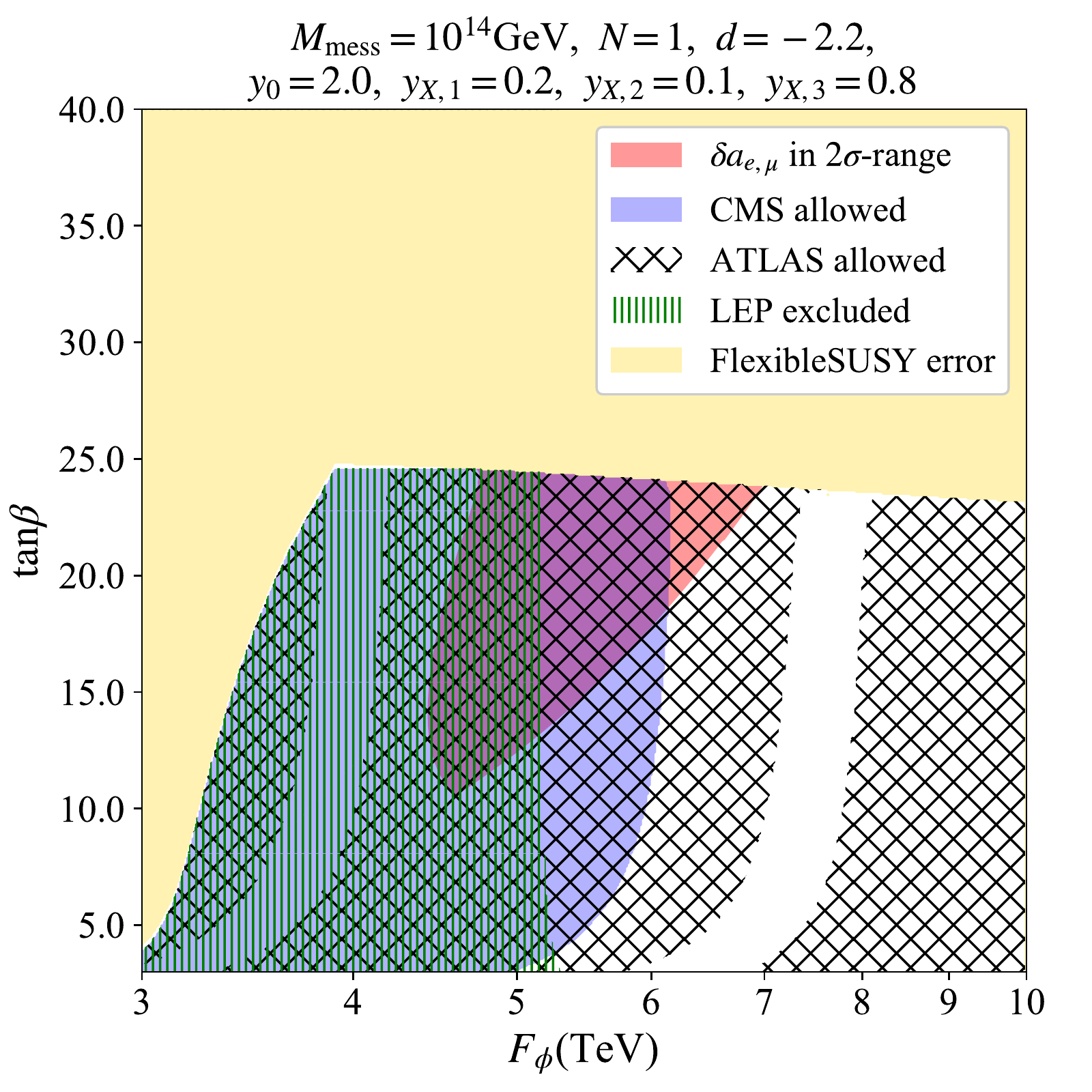}
	\includegraphics[width=.45\textwidth]{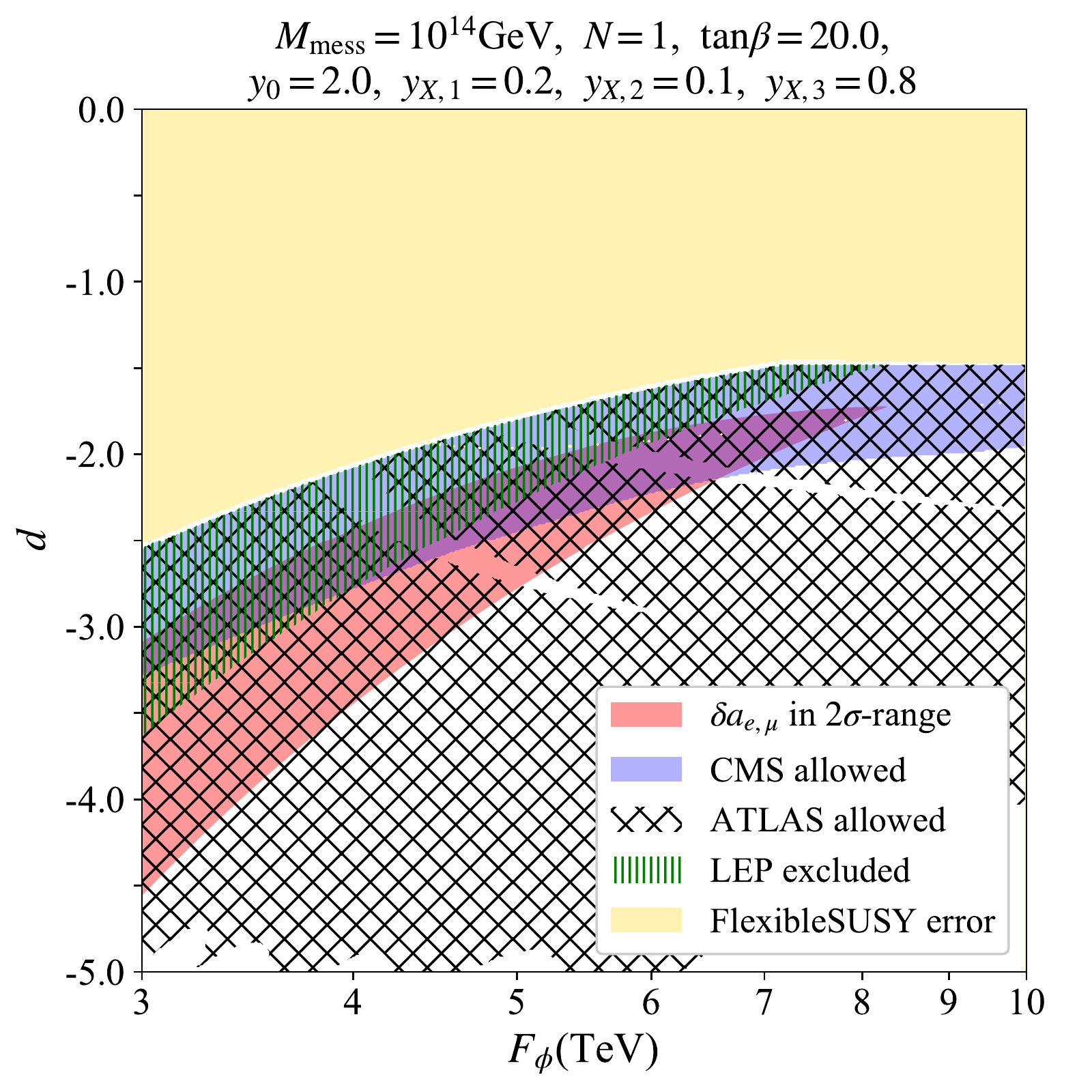}
	\caption{\label{fig:DAMSB-tanbeta-and-d-vs-fphi} 
	The electron and muon anomalous magnetic moments and collider constraints in the Yukawa defected AMSB with messenger-matter interactions, shown on the plane of $\tan\beta$ versus $F_\phi$ (left) and $d$ versus $F_\phi$ (right).} 
\end{figure*}

In this deflected AMSB scenario, we have $m_{\tilde\chi_1^\pm}\approx m_{\tilde\chi_2^0}$ in the parameter region that can explain both anomalous magnetic moments. The value of $\tan\beta$ cannot be too large, i.e., $\tan\beta\lesssim 25.0$  as shown in  Fig.~\ref{fig:DAMSB-tanbeta-and-d-vs-fphi}. 
If $\tan\beta$ is too large, it will cause $m_{h}$ (and $m_{A}$) to be tachyonic. Since the SUSY contribution to electron/muon $g-2$ is proportional to $\tan\beta$, if $\tan\beta$ is too small, the $g-2$ will not meet the requirement unless we fine-tune other parameters. The lightest neutralino mass $m_{\tilde\chi_1^0}$ will decrease as $F_\phi$ increases for fixed values of other parameters, which will be constrained by the CMS bounds. On the other hand, if $F_\phi$ is too small, it will also cause serious phenomenological difficulties, such as too light sleptons with masses less than 100 GeV. The value of the deflection parameter $d$ is directly related to the masses of wino and bino, so it will be constrained by the electron/muon $g-2$.  At the same time, since $d$ is related to the mass of $\tilde{\chi}_1^0$, it will also be stringently constrained by CMS searches. Further, in order to interpret $\delta a^{\rm SUSY}_\mu$, the right-handed smuon mass needs to get a large negative loop correction to reduce its value, and thus the LEP bounds will also exclude some portion of the parameter space for $d$. These constraints drive $d$ in a very narrow range. However, as we will see later, $y_{X,2}$ will be subject to even stringent constraints than that of $d$. There is a clear correlation between $d$ and $F_\phi$. This correlation can be seen from Eq.~\eqref{eq:DAMSB-binowinosoftmass}. To get enough SUSY contributions to both anomalous magnetic moments, the absolute values of $M_1$ and $M_2$ are also constrained to lie in a very narrow range. When $d$ decreases, $b_i-d\tilde{\Delta}$ will increase, and thus $F_\phi$ needs also to decrease.

\begin{figure*}[tbp]
	\centering 
	\includegraphics[width=.6\textwidth]{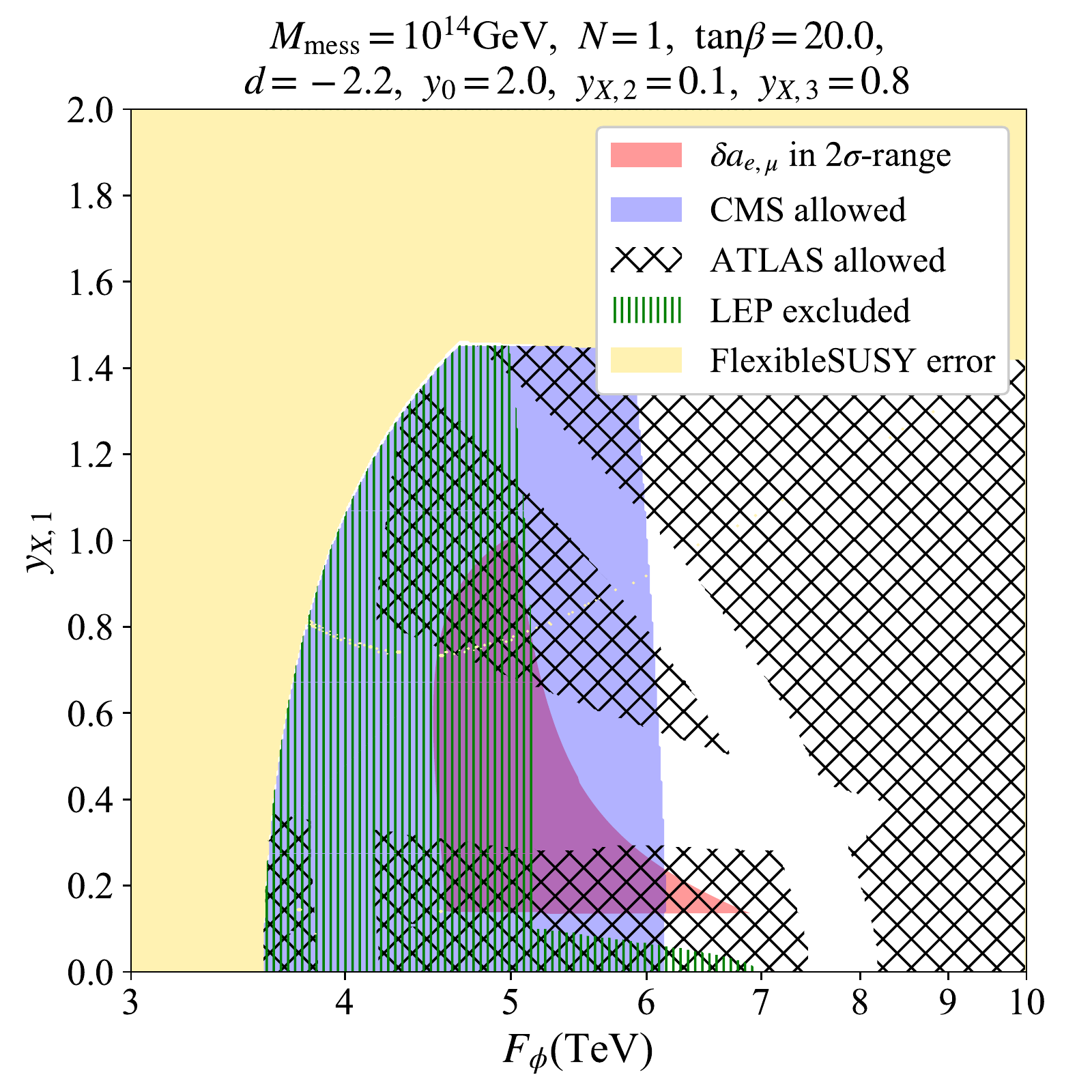}
	\caption{\label{fig:DAMSB-yx1-vs-fphi} 
		The electron and muon anomalous magnetic moments and collider constraints in the Yukawa defected AMSB with messenger-matter interactions, shown on the plane of $y_{X,1}$ versus $F_\phi$.} 
\end{figure*}

As shown in Fig.\ref{fig:DAMSB-yx1-vs-fphi}, the parameter $y_{X,1}$ is similarly restricted.  As $\delta a^{\rm SUSY}_\mu$ is sensitive to the right-handed smuon mass that needs a negative loop correction,  it is then indirectly sensitive to almost all relevant parameters. Of course, the dependence of the right-handed smuon mass on $y_{E,1}$ is relatively weak, and the value of $y_{E,1}$ can range from 0.1 to 1.0 when the collider bounds are not included. When $y_{E,1}$ is close to 0, the masses $m_{\tilde{\chi}_1^0}$ and $m_{\tilde{\chi}_2^0}$ are very close to each other. However, when $y_{E,1}$ gradually increases, the growth rate of $m_{\tilde{\chi}_1^0}$ will be larger than that of $m_{\tilde{\chi}_2^0}$, which thus will cause $\Delta(m_{\tilde{\chi}_1^0},m_{\tilde{\chi}_2^0})$ to increase. Finally, when $y_{X,1}\gtrsim 0.3$, such values will be excluded by ATLAS exclusion bounds. Unlike $d$, varying $y_{X,1}$ will not significantly affect the allowed range of $F_\phi$, which is similar to the effect of $\tan\beta$ on $F_\phi$.

\begin{figure*}[tbp]
	\centering 
	\includegraphics[width=.45\textwidth]{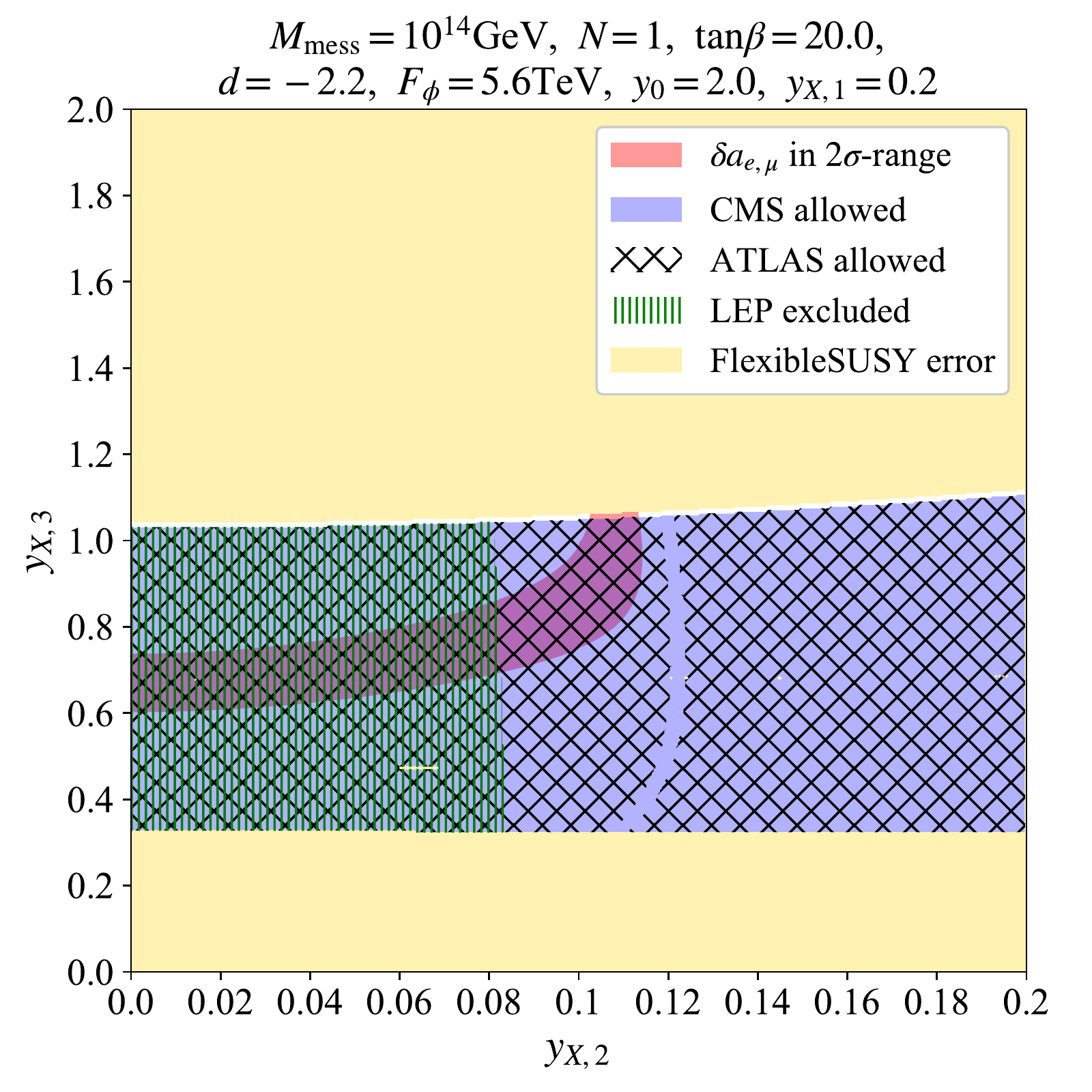}
	\includegraphics[width=.45\textwidth]{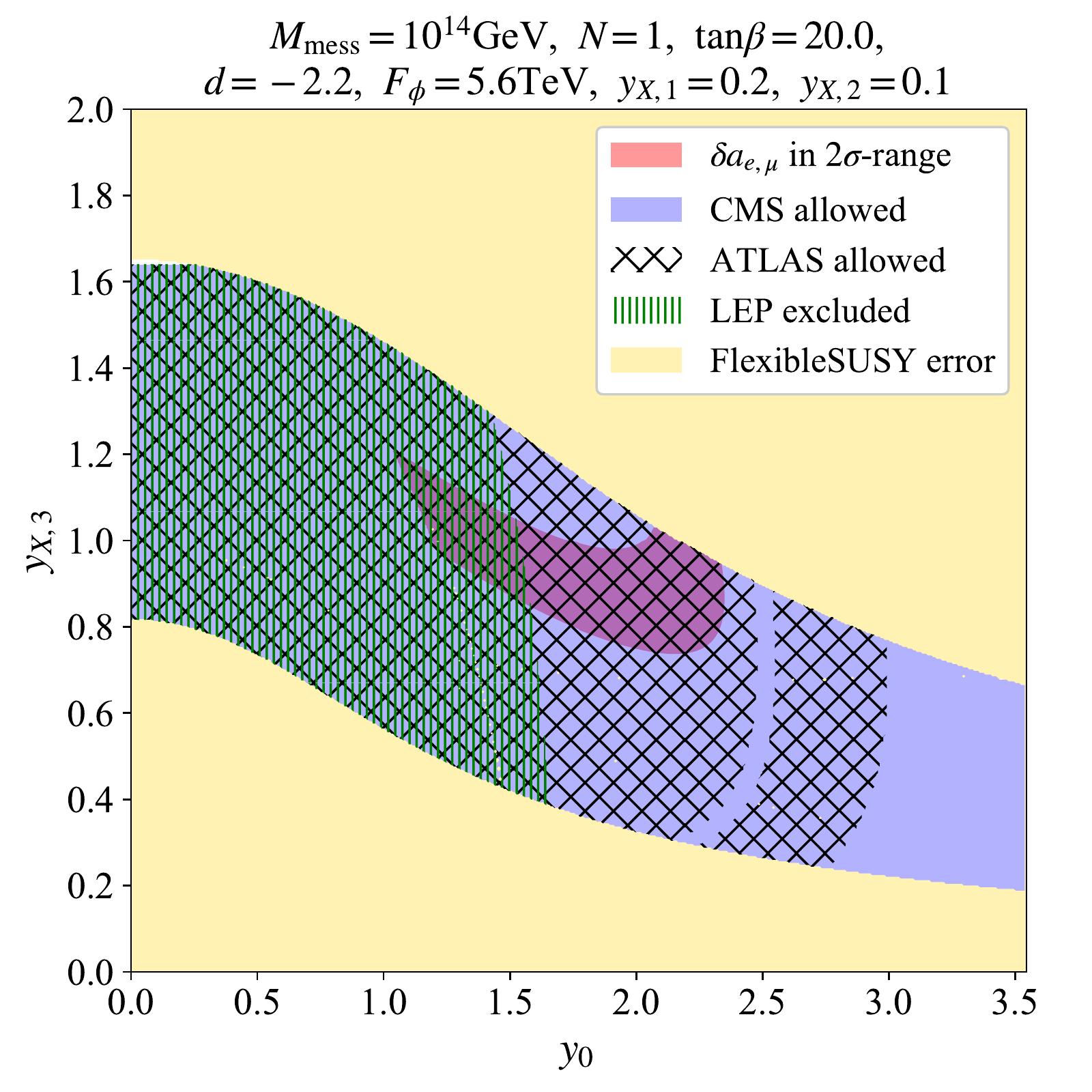}
	\caption{\label{fig:DAMSB-yukawa} 
	The electron and muon anomalous magnetic moments and collider constraints in the Yukawa defected AMSB with messenger-matter interactions, shown on the plane of $y_{X,3}$ versus $y_{X,2}$ (left) and $y_{X,3}$ versus $y_0$ (right).} 
\end{figure*}

Fig.~\ref{fig:DAMSB-yukawa} shows the allowed ranges of the Yukawa couplings in this deflected AMSB. Since $y_{X,2}$ is directly related to the right-handed smuon mass,  its value is restricted to a very narrow range below 0.12 to satisfy the requirements of muon anomalous magnetic moment. At the same time, we cannot simply make $y_{E,2}=0$, which will make the smuons too light to be excluded by the LEP bounds. Overall, we have $0.08\lesssim y_{E,2} \lesssim 0.12$. In this case, $y_{E,2}$ will be about one-tenth of $y_{E,3}$. 
The LEP bound on the stau mass is not shown in this figure. If the  LEP bounds on stau masses are taken into account, it will affect only the case with $y_{E,3}\lesssim 0.4$, and the region that can explain the anomalous magnetic moments will not change. When $y_{E,3}$ is too large, it will invalidate the perturbation theory and cause FlexibleSUSY to report errors. 
For the coupling $y_0$, its allowed range is much larger. Even if various experimental constraints are included, its allowed range is $1.5\leq y_0\leq 2.5$, which is much larger than that of other Yukawa couplings. Moreover, unlike other Yukawa couplings, perturbation bounds on $y_0$ is fairly mild. When $y_{X,3}$ is small, $y_0$ can even take the value $\sqrt{4\pi}$, which,  however, is excluded by the ATLAS experiment and also does not meet the requirements of anomalous magnetic moments.

Finally we stress that in this work we focused on the joint explanation of electron and muon $g-2$ anomalies. For the single explanation of muon $g-2$, the MSSM can readily make it (see, e.g., \cite{Abdughani:2019wai,Cox:2018qyi,Kobakhidze:2016mdx,VanBeekveld:2021tgn,Baum:2021qzx,Yang:2022gvz} ) while the GUT-constrained models like mSUGRA and GMSB need to be extended. \cite{Akula:2013ioa,Wang:2015nra,Wang:2015rli,Wang:2017vxj,Wang:2018vrr,Han:2020exx,Li:2021pnt,Wang:2021bcx,Chakraborti:2021bmv}. 

\section{Conclusions}
\label{sec:conclusions} 

We propose to provide a joint explanation of electron/muon $g-2$  anomalies in two UV-completed SUSY models in the framework of the anomaly mediation of SUSY breaking (AMSB). We constructed two Yukawa deflected AMSB type models, one with lepton-specific interactions and the other one with messenger-matter interactions.  Both models were found to be able to realize such an explanation:
\begin{itemize}
\item[(i)]  In the Yukawa deflected AMSB model with lepton-specific interactions, the mass ratio $m_{\tilde{\mu}_L}/m_{\tilde{\mu}_R}$ can be larger than 6, and $\delta a^{\rm SUSY}_e$ can be in the entire $2\sigma$ range. In the regions that can explain both anomalies, the parameters $F_{\phi}$ and $\tilde{\mu}$ demonstrate a strongly linear correlation. 
\item[(ii)]  In the Yukawa deflected AMSB with messenger-matter interactions, the mass ratio $m_{\tilde{\mu}_L}/m_{\tilde{\mu}_R}\lesssim 4$, which poses challenges to explain the two anomalies simultaneously. Because the masses of selectrons are not small enough, it is difficult for $\delta a^{\rm SUSY}_e$ to enter the $1\sigma$ range. This model originally was expected to realize the parameter regions in \cite{Badziak:2019gaf}. However, in the region that can explain both anomalies, the corresponding slepton masses were too low to survive the LEP bounds. Therefore, this deflected AMSB can only provide the parameter region in \cite{Li:2021koa}.
\item[(iii)] In both models, the couplings $y_{E,2}$ and $y_{X,2}$ are stringently constrained. In particular, $y_{X,2}$ is constrained to lie in the range 0.08-0.12 in the second model. 
\end{itemize}

\addcontentsline{toc}{section}{Acknowledgments}
\acknowledgments
This work was supported by the National Natural Science Foundation of China 
(NNSFC) under grant Nos. 12075213,  11821505 and 12075300,  
by the Key Research Project of Henan Education Department for colleges and universities under grant number 21A140025,
by Peng-Huan-Wu Theoretical Physics Innovation Center (12047503),
by the CAS Center for Excellence in Particle Physics (CCEPP), 
by a Key R\&D Program of Ministry
of Science and Technology of China under number 2017YFA0402204, and by the Key Research Program of the Chinese Academy of Sciences, Grant NO. XDPB15.

\addcontentsline{toc}{section}{References}
\bibliographystyle{JHEP}
\bibliography{bibliography}

\end{document}